\documentclass[reprint,amsmath,amssymb,aps,prl,twocolumn]{revtex4-2}
\pdfoutput=1
\usepackage{graphicx,amsmath,amssymb,amsthm}
\usepackage[hidelinks]{hyperref}
\usepackage{xcolor}
\usepackage{thmtools, thm-restate}
\usepackage{epsfig}
\usepackage{epstopdf}
\usepackage{latexsym}
\usepackage{booktabs}
\usepackage{dsfont}
\usepackage{bbm}
\usepackage{color}
\usepackage{physics}
\usepackage{tensor}
\usepackage{verbatim}
\usepackage[caption=false]{subfig}
\usepackage{tikz}
\usepackage{bigints}
\usepackage{ifthen}
\usetikzlibrary{matrix}
\usetikzlibrary{decorations.markings,calc,shapes,decorations.pathmorphing,decorations.pathreplacing,calligraphy}
\usetikzlibrary{patterns}
\usetikzlibrary{positioning}
\usepackage{textpos}

\usepackage{hyperref}

\usepackage{xcolor}

\hypersetup{
colorlinks,
linkcolor={blue!80!black},
citecolor={blue!80!black},
urlcolor={blue!80!black},
linktoc=page
}

\newcommand\beq{\begin{equation}}
\newcommand\eeq{\end{equation}}

\newcommand{\be}{\begin{equation}}
\newcommand{\ee}{\end{equation}}

\DeclareMathSizes{10}{10}{7}{7}

\begin{document}

\title{Scrambling in quantum cellular automata}

\author{Brian Kent}
\email{brian\_kent@utexas.edu}
\author{Sarah Racz}
\email{racz.sarah@utexas.edu}
\author{Sanjit Shashi}
\email{sshashi@utexas.edu}
\affiliation{Theory Group, Weinberg Institute, Department of Physics, University of Texas, 2515 Speedway, Austin, Texas 78712, USA.}

\preprint{\today}
\begin{abstract}
\noindent Scrambling is the delocalization of quantum information over a many-body system and underlies all quantum-chaotic dynamics. We employ discrete quantum cellular automata as classically simulable toy models of scrambling. We observe that these automata break ergodicity, i.e. they exhibit quantum scarring. We also find that the time-scale of scrambling rises with the local Hilbert-space dimension and obeys a specific combinatorial pattern. We then show that scarring is mostly suppressed in a semiclassical limit, demonstrating that semiclassical-chaotic systems are more ergodic.
\end{abstract}

\maketitle


\paragraph{\textbf{Introduction}.}
\textit{Scrambling} is the process of initially local information spreading out over a many-body system and ultimately becoming delocalized over time \cite{Hayden:2007cs,Sekino:2008he,Shenker:2013pqa}. It is the mechanism through which a quantum-chaotic system can achieve \textit{thermalization}, i.e. its relaxation to an ergodic state with no memory of the initial local degrees of freedom \cite{Deutsch:1991eth,Srednicki:1994mfb,Tasaki:1997ran,Rigol:2007thm}. Even if the system does not thermalize and instead (perhaps weakly) breaks ergodicity after long times---a phenomenon known as \textit{quantum scarring} \cite{Heller:1984zz,Turner2017WeakEB}---the information is still scrambled.

Scrambling underlies quantum-chaotic dynamics, so it has been the subject of serious investigation over the past several years in the study of quantum information, condensed matter theory, and quantum gravity \cite{Maldacena:2015waa,Faulkner:2022mlp}. More specifically, scrambling has provided insights into properties of computational complexity \cite{Susskind:2014moa,Stanford:2014jda,Belin:2021bga}, entanglement dynamics \cite{Hosur:2015ylk,Nahum:2017yvy}, and transport \cite{Blake:2016wvh,Davison:2016ngz,Hartman:2017hhp} in quantum systems. As black holes \cite{Hayden:2007cs,Sekino:2008he,Shenker:2013pqa} and other spacetimes \cite{Susskind:2011ap} can be thought of as ``fast" scramblers \cite{Lashkari:2011yi}, scrambling is also of interest in quantum gravity. Furthermore, there are protocols with which to measure scrambling through simulation and experiment \cite{Swingle:2016var,Xu:2018xfz,Xu:2022vko,Li:2016xhw,Zhu:2021uzs}. So, by studying scrambling, we glean profound, testable insight into the universal features of quantum dynamics.

Tunable toy models that highlight phenomena of interest are a tried and true tool of theoretical physics. In this paper, we propose that \textit{quantum cellular automata} (QCAs) \cite{Lent:1993qca,Schumacher:2004qca,Farrelly:2019zds} are a useful tool for studying scrambling (cf. \cite{Gopalkrishnan_2018,Gopalakrishnan_20182,Iadecola_2020,Hillberry:2020nfj,Sellapillay:2022jkg,Farshi:2022clf,Sommers:2022cuz}). These are lattice systems equipped with local Hilbert spaces on each site and a discrete time-evolution operation. This operation can be phrased in the language of Heisenberg (time-dependent) operators, and so it is straightforward to simulate operator growth from some initial state. Thus we can apply the general protocol for studying scrambling \cite{Swingle:2016var,Xu:2018xfz,Xu:2022vko} to QCAs.

To exemplify the utility of cellular automata, we elucidate the scrambling behavior of a particular class: ``Clifford" QCAs. We find that they harbor scarring for particular initial conditions. We also simulate the dependence of the ``scrambling time" on the Hilbert-space dimension $N$, with large $N$ describing a semiclassical regime. We ultimately find that the semiclassical systems are ``more" ergodic than low-$N$ ones at late times.

\begin{figure}
\centering
\begin{tikzpicture}[scale=0.9]
\node at (-2,0) {};
\draw[pattern=north west lines,draw=none] (0,1) to (2,3) to (-2,3) -- cycle;

\draw[-,dashed,thick] (0,0) to (2,0);
\draw[-,thick] (0,0) to (0,1);
\draw[-,thick] (2,0) to (2,3);

\draw[-,thick,dashed] (0,1) to (2,3);
\draw[-,thick,dashed] (0,1) to (-2,3);

\draw[->] (-1.7,0.85+0.15) to (-1.7,2+0.15);
\draw[->] (-1.85,1+0.15) to (-0.7,1+0.15);

\draw [decorate, thick, decoration = {brace,raise=2pt}] (0,0.05) --  (0,0.95);
\node at (-0.35,0.5) {$t_*$};

\draw [decorate, thick, decoration = {brace,raise=2pt}] (2,2.95) --  (2,0.05);
\node at (2.35,1.5) {$t$};
\node[white] at (-2.35,1.5) {$t$};

\draw [decorate, thick, decoration = {brace,raise=2pt}] (2,0) --  (0,0);
\node at (1,-0.35) {$v_{\text{B}}(t-t_*)$};

\node at (-1.95,2+0.15) {$t$};
\node at (-0.7,0.75+0.15) {$x$};

\node[rotate=45] at (1+0.17-0.08,2-0.17-0.08) {$C(x,t) \sim O(1)$};

\node at (0,0) {$\bullet$};
\draw[->] (0,0) to[bend left] (-1,0);

\node at (-1.15,0.05) {$V$};

\node at (2,0) {$\bullet$};
\draw[->] (2,0) to[bend right] (3,0);

\node at (3.5,0.05) {$W(x)$};
\node[white] at (-3.5,0.05) {$W(x)$};

\end{tikzpicture}
\caption{A schematic ``space-time" plot depicting scrambling between initial operators $V$ and $W(x)$. For a given $x$, the squared commutator eventually reaches $O(1)$. Over all $x$, this defines a cone (the shaded region) in which information is diagnosed as scrambled. The minimum time $t_*$ of this cone is the scrambling time. The inverse slope $v_{\text{B}}$ is the associated velocity.}
\label{figs:butterflyCone}
\end{figure}
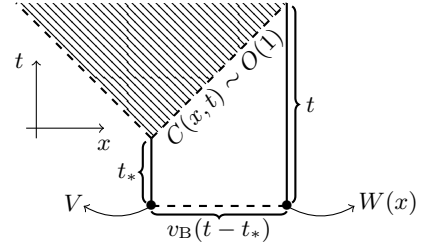

\paragraph{\textbf{Quantifying scrambling.}} We quantify scrambling with the \textit{out-of-time-ordered correlator} (OTOC) \cite{Swingle:2016var,Xu:2018xfz,Xu:2022vko},
\begin{equation}
F(x,t) = \expval{W(x,t)^\dagger V^\dagger W(x,t) V}.
\end{equation}
$V$ is a local operator. $W(x,t) = U(-t)W(x)U(t)$ is the time evolution of a local operator $W(x)$ inserted at position $x$. We take $[V,W(x)] = 0$, so the OTOC measures the breakdown of commutativity due to operator growth.

The OTOC is used to study features of chaos, such as operator growth and the butterfly effect. To see why, we write it in terms of the ``squared commutator"
\begin{equation}
\begin{split}
C(x,t)
&= \expval{[W(x,t),V]^\dagger [W(x,t),V]}\\
&= 2\left[1- \text{Re}\,F(x,t)\right],
\end{split}
\end{equation}
assuming unitary operators. We call the system ``scrambled" when this quantity is $O(1)$. For local interactions, the growth of $C(x,t)$ is said to take a universal form \cite{Xu:2018xfz}:
\begin{equation}
C(x,t) \sim \mathcal{C}\exp\left\{-\frac{\lambda_{\text{L}}}{t^{p}}\left(\frac{|x|}{v_{\text{B}}}-t\right)^{1+p} \right\}.\label{boundComm}
\end{equation}
$\mathcal{C}$, $p$, $\lambda_{\text{L}}$, $v_{\text{B}}$ (the \textit{butterfly velocity}) are constants. In \eqref{boundComm}, $C(x,t)$ is not $O(1)$ until $t \sim |x|/v_{\text{B}}$. By depicting the squared commutator on a ``space-time" ($t$ vs. $x$) plot, we observe a ``butterfly cone" of velocity $v_{\text{B}}$ beyond which we diagnose the system as scrambled (e.g. Figure \ref{figs:butterflyCone}) \footnote{{\eqref{boundComm} and its bound in the holographic limit $p \to 0$ \cite{Roberts:2016wdl} are state-dependent, i.e. $\lambda_{\text{L}}$ and $v_{\text{B}}$ care about $V$ and $W(x)$. There is a similar, state-independent bound of this form with $p = 0$ called the \textit{Lieb--Robinson bound} on the microscopic norm (as opposed to the average) of the squared commutator \cite{Lieb:1972wy,Nachtergaele_2006,Hastings_2006,Hastings:2010lrb}. \cite{Swingle:2016var} interprets the butterfly cone as a low-energy effective ``Lieb--Robinson" cone.}}. The minimum time $t_*$ of this cone is the \textit{scrambling time}.

\paragraph{\textbf{Clifford QCAs}.}

QCAs are lattice systems that undergo discrete time evolution. Each is determined by two things: the local Hilbert spaces on each lattice site and the unitary time-evolution operator (or automorphism). In the Heisenberg picture, we may write the latter as a set of reversible ``rules" \cite{Schumacher:2004qca} for how local operators on each site evolve.

We consider a particular class of model systems called \textit{Clifford quantum cellular automata} \cite{Schlingemann_2008,Gutschow_2009,Gutschow_2010,Berenstein:2018zif}. These QCAs live on an infinite 1d lattice in space and obey translation invariance. The Hilbert space of each lattice site arises from quantizing a toroidal phase space, so each local Hilbert space is finite-dimensional \cite{2003AmJPh..71...49D}. We denote this dimension as $N$. Furthermore, the Planck constant scales as $1/N$ \cite{Berenstein:2018zif}, and so $N \to \infty$ is a semiclassical limit.

The operators acting on each local Hilbert space constitute a generalized Clifford algebra generated by $Q,P$:
\begin{equation}
Q^N = P^N = \mathds{1},\ \ P Q = \omega Q P.\label{clifford}
\end{equation}
$\mathds{1}$ is the identity and $\omega = e^{2\pi i/N}$. We write the generators in Sylvester's $N\times N$ ``clock-and-shift" representation:
\begin{equation}
Q = \begin{pmatrix}
1 & 0 & \cdots & 0\\
0 & \omega & \cdots & 0\\
\vdots & \vdots & \ddots & \vdots\\
0 & 0 & \cdots & \omega^{N-1}
\end{pmatrix},\ \ P = \begin{pmatrix}
0 & 1 & 0 & \cdots & 0\\
0 & 0 & 1 & \cdots & 0\\
\vdots & \vdots & \vdots & \ddots & \vdots\\
0 & 0 & 0 & \cdots & 1\\
1 & 0 & 0 & \cdots & 0
\end{pmatrix}.
\end{equation}
These matrices and their products (excluding $\mathds{1}$) are also called \textit{generalized Pauli operators}. Indeed, taking $N = 2$ reproduces the usual Pauli matrices that generate $\mathfrak{su}(2)$.

Next, while there are many rules one may implement (so long as they are reversible and translation invariant along the lattice), for specificity we will focus on
\begin{equation}
\begin{split}
Q_\alpha &\to Q_{\alpha-1} \otimes Q_\alpha P_\alpha \otimes Q_{\alpha+1},\\
P_\alpha &\to (Q_\alpha)^{N-1},
\end{split}\quad \forall \alpha \in \mathbb{Z}.\label{ourRule}
\end{equation}
The index $\alpha$ denotes the lattice position. We describe the time evolution of any operator on the lattice by iteratively implementing these (nearest-neighbor) rules. 

The same machinery used for probing scrambling can be used for Clifford QCAs. In fact, the squared commutator's form can be computed analytically if we have reflection symmetry about $\alpha = 0$. If we take
\begin{equation}
\begin{split}
V &= \cdots \otimes \mathds{1}_{-1} \otimes \Sigma_0 \otimes \mathds{1}_1 \otimes \cdots,\\
W_\alpha &= \cdots \otimes \mathds{1}_{\alpha-1} \otimes \widetilde{\Sigma}_{\alpha} \otimes \mathds{1}_{\alpha+1} \otimes \cdots,
\end{split}\label{initOps}
\end{equation}
where $\Sigma_0$ and $\widetilde{\Sigma}_\alpha$ are possibly distinct generalized Pauli matrices respectively at sites $0$ and $\alpha$, then the squared commutator has the form (cf. supplemental material)
\begin{equation}
\begin{split}
C_\alpha(t)
&= \expval{[W_\alpha(t),V]^\dagger [W_\alpha(t),V]}\\
&= 4\sin^2\left[\frac{\pi}{N}\xi(\alpha,t)\right],\label{doubleCommCQCA}
\end{split}
\end{equation}
where $\xi(\alpha,t)$ is some integer function of space-time.

At first glance, this oscillatory behavior may appear troubling for our claim that Clifford QCAs exhibit scrambling. We do not get something of the form \eqref{boundComm}, but this is how we argue for a butterfly cone. Not having \eqref{boundComm} may prevent us from diagnosing the system as scrambled.

Fortunately, there is a loophole; we simply need a cone along whose boundary the squared commutator reaches $O(1)$. So long as we have such a cone, we do not need to care about the late-time dynamics. Indeed, when we simulate the evolution of the Clifford QCA with the rule \eqref{ourRule}, we will certainly find that sinusoidal behavior \eqref{doubleCommCQCA} persist in the late-time dynamics. Nonetheless, we will also consistently find the requisite cones from scrambling.

\paragraph{\textbf{Scrambling in action}.}

\begin{center}
\begin{figure}
\centering
\includegraphics[trim = {4.5cm 0 5cm 0.5cm},clip,scale=0.45]{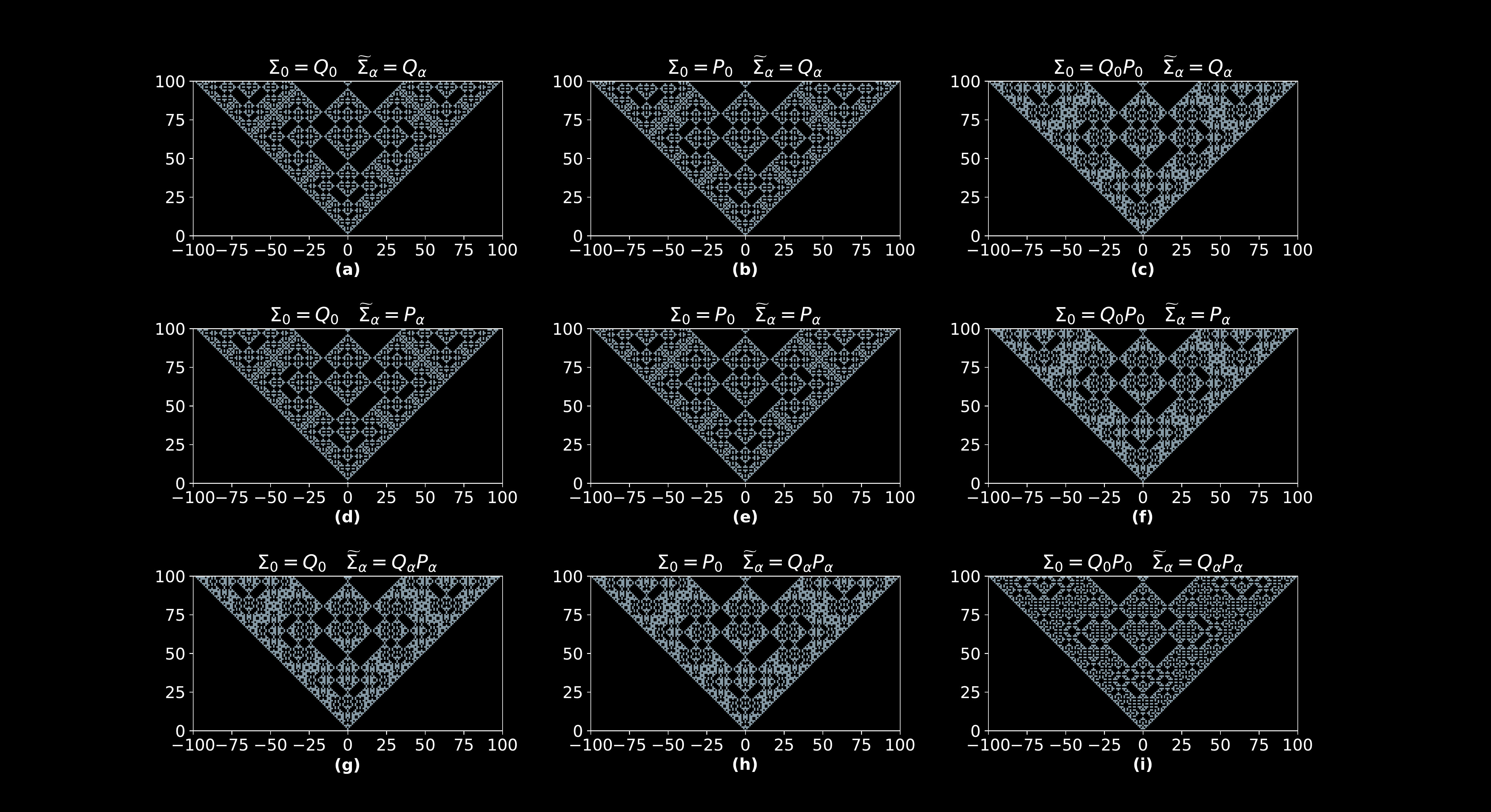}
\caption{All nine space-time ``heat" plots of the squared commutators $C_\alpha(t) = \expval{[W_\alpha(t),V]^\dagger [W_\alpha(t),V]}$ with local initial insertions \eqref{initOps} in the $N = 2$ Clifford QCA with rule \eqref{ourRule}. We restrict to lattice sites $-100 \leq \alpha \leq 100$ and $100$ time steps. Reading across columns, the initial operators at the central ($0$) lattice site are (a,d,g) $Q_0$, (b,e,h) $P_0$, and (c,f,i) $Q_0 P_0$. Reading across rows, the initial operators at the $\alpha$ lattice site are (a)--(c) $Q_\alpha$, (d)--(f) $P_\alpha$, and (g)--(i) $Q_\alpha P_\alpha$. $C_\alpha(t) = 4$ at blue points and $0$ at black points. Comparing all initializations, the small-scale dynamics subtly differ, but the large-scale dynamics yield fractals of similar sharpness. More precisely, we find that all patterns have approximately the same fractal dimension $\sim 1.83$ (see supplemental material for the calculation), consistent with the exact value of the ``trace-time" fractals\footnote{We thank Grace Sommers for pointing out that this exact value is known.} $\log_{2} \frac{3+\sqrt{17}}{2}$ \cite{Gtschow2010TheFS,Sommers:2022cuz}.}
\label{figs:spacetimeN2}
\end{figure}
\end{center}

There are $N^2 - 1$ generalized Pauli matrices acting on the Hilbert space of dimension $N$, so we have precisely $(N^2-1)^2$ squared commutators computed from local insertions. As an example, we present the nine space-time ``heat" plots for the simplest case $N = 2$ in Figure \ref{figs:spacetimeN2}. From the analytic form of the commutator \eqref{doubleCommCQCA}, $C_\alpha(t)$ is either $0$ or $4$, so loss of commutativity is akin to a bit flip.

In all of these plots, we observe a cone along which the squared commutator reaches $O(1)$. Thus, there is scrambling according to our diagnostic. However, the dynamics within the cone, particularly at late times, yield a fractal pattern because of the sinusoidal behavior \eqref{doubleCommCQCA}. Such behavior has also been observed in previous studies of Clifford QCAs plotting other quantities (cf. \cite{Gutschow_2009,Gutschow_2010,Gtschow2010TheFS,Berenstein:2021lya}), but our analysis highlights that these fractals arise from the chaotic dynamics after scrambling.

We claim this to be a violation of ergodicity, signaling quantum scarring. To see why, first consider how we would diagnose a system as thermalized at late times. Thermalization of the system means a loss of ``memory" of the initial insertions. Such memory must instead be confined to subleading effects in the late-time regime deep in the cone. So, if the late-time dynamics are ergodic, then the squared commutator inside of the cone must lose track of the separation between the initial insertions. In other words, it must exhibit $x$ independence.

That is not what we see in our Clifford QCA. Instead, the sinusoidal behavior \eqref{doubleCommCQCA} persists at late times and induces the fractal pattern seen in our space-time plots. So, although the information is scrambled (i.e. delocalized) within the cone, it is not thermalized (i.e. randomized) over space. The fractals represent quantum scarring.

In this paper, our focus is on OTOCs of local initial operators (cf. \cite{Roberts:2014isa}). However, the OTOC can generally probe the loss of commutativity between initially commuting \textit{nonlocal} operators, as well. The initial operators themselves need not be local. For instance, we may start with an initial operator $W_\alpha$ which comprises generalized Pauli matrices localized to multiple lattice sites. It would be interesting to simulate such initializations to understand the dynamics of nonlocal states.
\pagebreak

\onecolumngrid
\vspace{-1.015cm}
\twocolumngrid

\onecolumngrid
\begin{center}
\begin{figure}
\centering
\includegraphics[trim = {4cm 1.5cm 7.75cm 2.5cm},clip,scale=0.45]{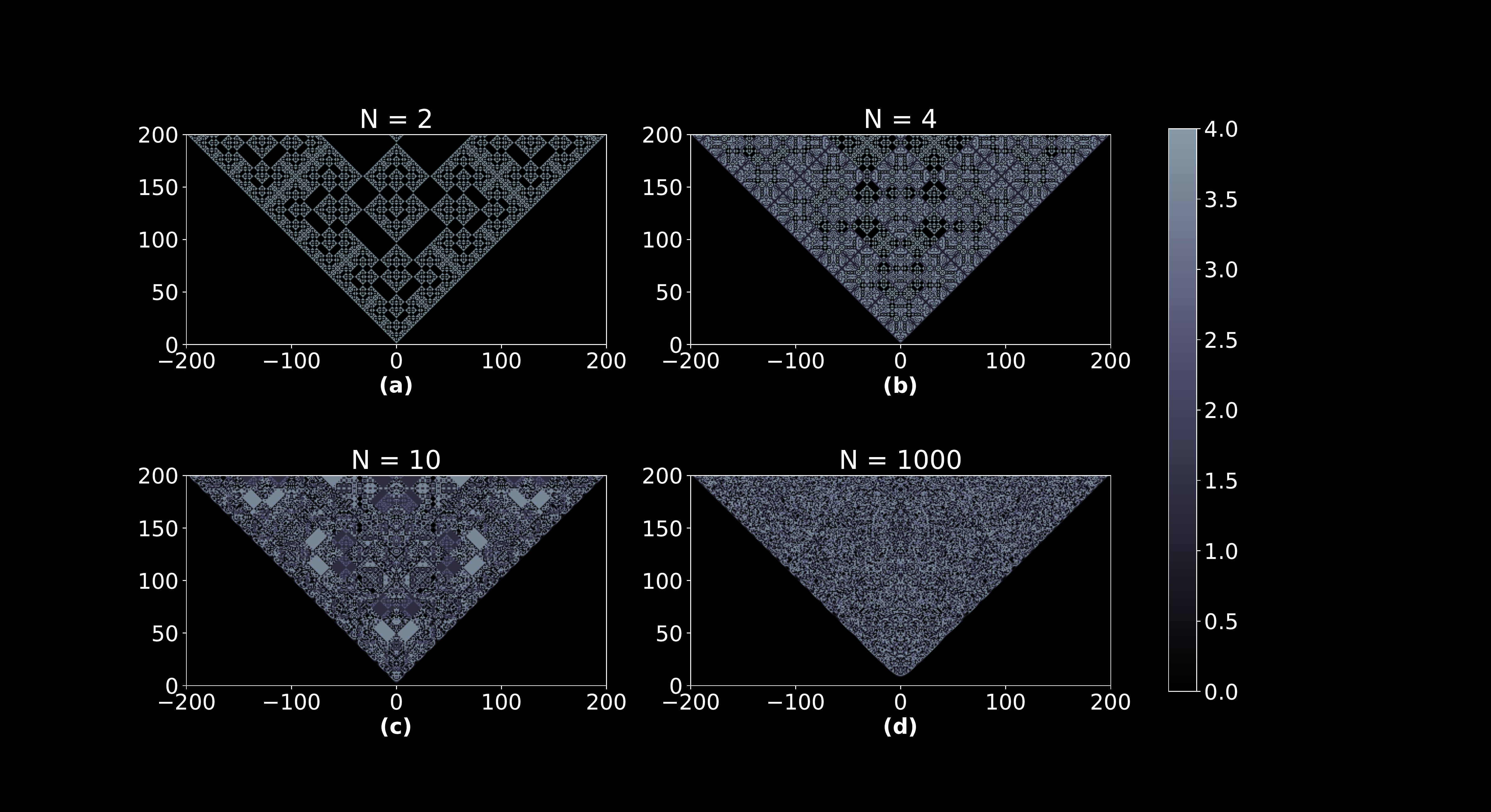}
\caption{The space-time heat plots depicting the squared commutators for initializations $\Sigma_0 = Q_0$, $\widetilde{\Sigma}_{\alpha} = Q_\alpha$ and (a) $N = 2$, (b) $N = 4$, (c) $N = 10$, and (d) $N = 1000$. We restrict to lattice sites $-200 \leq \alpha \leq 200$ and $200$ time steps. Unlike in Figure \ref{figs:spacetimeN2}, we must use a gradient to represent the values of $C_\alpha(t)$ since we simulate $N > 2$. Larger $N$ yields a later cone and, thus, larger scrambling time. Furthermore, the fractal pattern is strongest for $N = 2$ and is ``filled in" for larger $N$. By $N = 1000$, the squared commutator has lost its position dependence deep inside of the cone, and so the late-time dynamics look more ergodic.}
\label{figs:dcHeatMaps}
\end{figure}
\end{center}
\vspace{-1.015cm}
\twocolumngrid

\paragraph{\textbf{Semiclassical regime}.}

Quantum chaos can be quite different from classical chaos. For example, classically chaotic systems always thermalize at late times, but quantum systems may scar. To reconcile the classical and quantum regimes, we may consider a semiclassical limit of a quantum chaotic system. In our Clifford QCAs, we can implement such a limit by increasing $N$. Specifically, we examine the influence of $N$ on the evolution of the initial local operator insertions $\Sigma_0 = Q_0$ and $\widetilde{\Sigma}_{\alpha} = Q_{\alpha}$. The space-time plots are presented in Figure \ref{figs:dcHeatMaps}.

We first examine the scrambling time $t_*$. It can be read off explicitly from the position of the emergent butterfly cone. By scanning over a range of $N \geq 2$, we find that $t_*$ ``jumps" by $1$ at particular values of $N$ (Table \ref{table:scramblingTimes}). Up to a factor of $6$, these values of $N$ equate to a subsequence of particular \textit{Whitney numbers} \cite{intseq} (defined later):
\begin{equation}
W_{2t} =\,\!_4F_3\left(\frac{1-t}{2},\frac{1-t}{2},-\frac{t}{2},-\frac{t}{2};1,-t,-t;16\right).\label{whitneyOurs}
\end{equation}
This specific sequence originates from our initialization ($\Sigma_0 = Q_0$, $\widetilde{\Sigma}_{\alpha} = Q_\alpha$) and our rule \eqref{ourRule}. To see why, note that we are simply finding the minimum time $t_*$ at which
\begin{equation}
C_\alpha(t_*) \geq 1 \implies \xi(0,t_*) \geq \frac{N}{6},\label{ineq}
\end{equation}
with fixed $N$ [recall \eqref{doubleCommCQCA}]. For \eqref{ourRule}, we find $\xi(0,t) = W_{2t}$.

\begin{table}[b]
\centering
\begin{tabular}{c||c|c|c|c|c|c c}
$N$ & $[2,6]$ & $[7,12]$ & $[13,30]$ & $[31,66]$ & $[67,156]$ & $[157,378]$ & $\cdots$\\\hline
$t_*$ & 1 & 2 & 3 & 4 & 5 & 6 & $\cdots$\\\hline
$\xi(0,t_*)$ & 1 & 2 & 5 & 11 & 26 & 63 & $\cdots$
\end{tabular}
\caption{A table listing scrambling times $t_*$ for different ranges of Hilbert-space dimensions $N$. As we increase $N$, $t_*$ ``jumps" at particular values of $N$. This is because the values of $\xi(0,t_*)$ appearing in the analytic form \eqref{doubleCommCQCA} of the squared commutator $C_0(t)$ constitute a particular rule-dependent and monotonic integer sequence \eqref{whitneyOurs}. For each range of $N$, there is a minimum $\xi(0,t_*)$ for which the squared commutator is $\geq 1$.}
\label{table:scramblingTimes}
\end{table}

So, consider some $N$ and the minimum $t_*$ for which \eqref{ineq} is satisfied. By increasing $N$, we will eventually violate this bound, and so we must go to the Whitney number for $t_* + 1$ for the squared commutator to be $\geq 1$.

This sequence of Whitney numbers appears in combinatorial graph theory \cite{Munarini2002OnTR,Conflitti2005OnWN} as follows. Define a \textit{fence} of order $n$ as a set of points $\{p_1,...,p_{n}\}$ imbued with partial ordering $p_1 < p_2 > p_3 < p_4 > \cdots$. An \textit{ideal} (with respect to the partial ordering) of order $i$ is any size-$i$ subset $S$ with the property that $p_k < p_j$ for any $p_j \in S$ implies $p_k \in S$. The Whitney number $f_{n,i}$ is then the number of order-$i$ ideals in the fence of order $n$ \footnote{{We can also define Whitney numbers for other partially ordered sets, such as ``crowns," which are cycilic fences \cite{Munarini2002OnTR,Conflitti2005OnWN}.}}.

Our Whitney numbers \eqref{whitneyOurs} count the ideals of order $t$ in the fence of order $2t$ and form a monotonic integer sequence indexed by $t$. This is only one possible sequence, however, and different initializations or rules may yield others. It would be interesting to see whether these sequences also have analogous combinatorial-graph-theoretic interpretations. We leave further exploration of this connection to future work.

We now briefly touch on the effect of increasing $N$ on scarring. It is expected that scars will be suppressed in the semiclassical limit, since we should recover ergodicity in this regime. This is precisely what we see in Figure \ref{figs:dcHeatMaps}. As we increase $N$, the squared commutator remains $O(1)$ within ``more" of the cone. This makes analytic sense in \eqref{doubleCommCQCA}; the period of the oscillations goes as $N$, so it takes longer for the squared commutator to fall back to $0$.

There is one caveat to this lesson even in large-$N$ systems. By an argument of Schwinger \cite{SchwingerBook}, the generalized Clifford algebra contains commuting Clifford subalgebras corresponding to distinct prime factors of $N$. If $N$ is composite, then these are proper subalgebras. These allow for what we call ``primal scars" when examining the dynamics of particular operator insertions for composite $N$.

For concreteness, suppose that $N = \kappa p^\ell$ for $p$ prime, $\kappa$ coprime to $p$, and some $\ell \geq 1$. Then, consider the subalgebra generated by the set $\{Q^\kappa,P^\kappa\}$. By \eqref{clifford},
\begin{align}
&\left(Q^\kappa\right)^{p^\ell} = \left(P^\kappa\right)^{p^\ell} = \mathds{1},\\
&P^\kappa Q^\kappa  = \omega^{\kappa^2} Q^\kappa P^\kappa = e^{2\pi i \kappa/p^\ell} Q^\kappa P^\kappa.
\end{align}
As $\kappa$ is coprime to $p$ (and thus $p^\ell$), $e^{2\pi i \kappa/p^\ell}$ is a primitive $p^\ell$-th root of unity. Thus, $\{Q^\kappa,P^\kappa\}$ generates a Clifford subalgebra that acts on the Hilbert space of size $p^\ell$. 

\begin{figure}
\centering
\includegraphics[trim = {2.75cm 1.25cm 3cm 2cm},clip,scale=0.2]{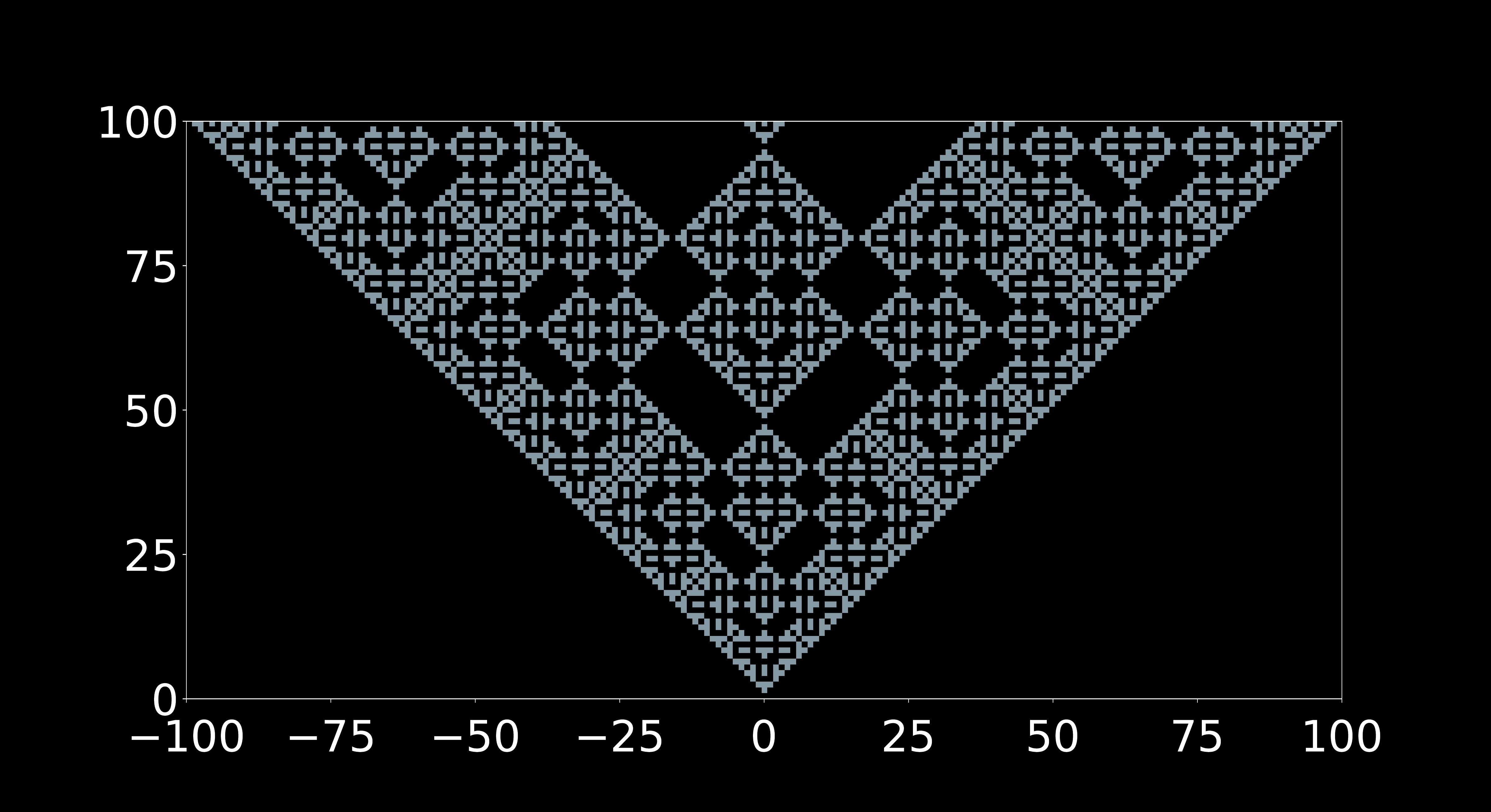}
\caption{The space-time heat plot of the squared commutator for $\Sigma_0 = Q_0^5$, $\widetilde{\Sigma}_{\alpha} = Q_\alpha^5$ with $N = 10$. Observe that this precisely matches one of the $N = 2$ plots---specifically Figure \ref{figs:spacetimeN2}, top left---because $10 = 5 \times 2$, and so we have that $C_\alpha(t)$ is $4$ (blue) or $0$ (black). This pattern is a primal scar.}
\label{figs:primalScars}
\end{figure}

We observe that this subalgebra is closed under our rule \eqref{ourRule}. As such, the simulated dynamics from the initialization $\Sigma_0 = Q_0^\kappa$ and $\widetilde{\Sigma}_{\alpha} = Q_\alpha^\kappa$ with $N = \kappa p^\ell$ yields the exact same pattern as the initialization $\Sigma_0 = Q_0$ and $\widetilde{\Sigma}_{\alpha} = Q_\alpha$ with $N = p^\ell$. See for example Figure \ref{figs:primalScars}, in which we simulate $\Sigma_0 = Q_0^5$ and $\widetilde{\Sigma}_{\alpha} = Q_\alpha^5$ for $N = 10$ and obtain a pattern from the $N = 2$ case (Figure \ref{figs:spacetimeN2}, top left).

In general, each distinct prime factor $p$ of $N$ labels a family of primal scars so long as the Clifford subalgebra associated with $p$ is preserved under the rule of the QCA. \eqref{ourRule} is one example of such a rule, but there are others.

However, the existence of primal scars does not completely spoil ergodicity in large-$N$ systems. Firstly, the space-time plots of the squared commutator generated from $\Sigma_0 = Q_0$ and $\widetilde{\Sigma}_{\alpha} = Q_\alpha$ do not maintain this type of scarring as we increase $N$. Secondly, we can restrict ourselves purely to prime values of $N$. In this case, there are no Clifford subalgebras even for large $N$, and so we would not have primal scars in the first place.

\paragraph{\textbf{Conclusions}.}
To summarize, quantum cellular automata are classically simulable toy environments for quantum many-body physics. In this paper, we elaborate on how nontrivial aspects of the scrambling dynamics of many-body systems can be easily studied by using QCAs. By focusing on operator growth in Clifford QCAs both analytically and numerically, we access concrete data quantifying scrambling, particularly butterfly velocities and scrambling times. Notably, we observe the formation of quantum scars, a feature of many-body chaos unique to the quantum regime. This further validates QCAs as a good arena for studying scarring \cite{Gopalkrishnan_2018,Gopalakrishnan_20182,Iadecola_2020}.

We find a deep connection between the structure of our Clifford QCA and combinatorial graph theory. Furthermore, we are able to study how signals of quantum chaos change in the semiclassical limit. By increasing the size of the local Hilbert space (a large-$N$ limit), we find that the fractal behavior exhibited in the squared commutator is typically ``filled in," leading the region within the butterfly cone to appear more thermal. While there is still primal scarring in that the dynamics of particular initializations for large composite values of $N$ produce ``sharper" fractals, increasing $N$ nonetheless always yields successively more random patterns.

In this paper, we focus on one rudimentary probe of scrambling---the squared commutator, which is a cousin of the out-of-time-ordered correlator. However, one may also examine more refined probes of post-scrambling-time behavior. One candidate is Krylov complexity \cite{Rabinovici:2020ryf}, which is defined in terms of Heisenberg operator evolution and has been proposed as a probe of quantum scarring \cite{Bhattacharjee:2022qjw}. Since the rule of a QCA is typically straightforward in the Heisenberg picture, it is natural also to study Krylov complexity in such systems.

A major focus of this work is studying semiclassical (large-$N$) operator growth to bridge the quantum regime to classical physics. This has also been the motivation underlying previous work \cite{Liu_2021,Pizzi_2022}. While we only have asked what happens with regards to scarring, it would be interesting to see if the large-$N$ limit of our Clifford QCAs explicitly reproduce the features of classical information spreading.

There are a variety of ways to change the type of physics being simulated. One can change the rule (even making them random \cite{Richter:2022clf}), the local Hilbert space, or the structure of the lattice (cf. \cite{Farshi:2022clf,Sommers:2022cuz}), thereby implementing alternate types of many-body systems. We can include multiple species, asymmetries, nonlocal interactions, or even various types of boundary conditions (reflecting, periodic, etc.). Overall, the world of quantum cellular automata is a vast toybox in which to test a myriad of ideas about quantum many-body dynamics and chaos.

\section*{Acknowledgements}

We thank Elena Caceres, Jacques Distler, Logan Hillberry, and Shenglong Xu for useful discussions. We also thank Grace Sommers for comments. The authors acknowledge support from National Science Foundation (NSF) Grant No. PHY-1914679. BK and SS are also supported by NSF Grant No. PHY-2112725.

\vfill\pagebreak

\section{Supplemental Material}

\subsection{Implementation of QCAs}

To implement a QCA on a computer, we must employ a method that accommodates the infinitude of lattice sites by using only common software architecture. We can do so by representing particular operators as Laurent polynomials whose coefficients are integers taken modulo $N$, as in \cite{Gutschow_2009,Gutschow_2010}. Specifically, we map the full operator algebra to a two-dimensional vector space and multiplication in the former to addition in the latter. Under this mapping, we represent generators at site $\alpha$ as
\begin{equation}
Q_\alpha \longmapsto \begin{pmatrix}
q^\alpha\\
0
\end{pmatrix},\ \ P_\alpha \longmapsto \begin{pmatrix}
0\\
q^\alpha
\end{pmatrix},
\end{equation}
where $q$ is just an abstract variable. So, under this mapping we write a generic operator as
\begin{equation}
\bigotimes_{\alpha = -\infty}^{\infty} Q_\alpha^{i_\alpha} P_\alpha^{j_\alpha} \longmapsto \sum_{\alpha = -\infty}^\infty \begin{pmatrix}
i_\alpha q^\alpha\\
j_\alpha q^\alpha
\end{pmatrix},\label{vecgen}
\end{equation}
with $i_\alpha,j_\alpha$ taken mod $N$. In this notation, a time-evolved operator $\mathcal{O}(t)$ is written
\begin{equation}
\mathcal{O}(t) = \omega^{\varphi(t)} M^{t} \mathcal{O}(0)\ (\text{mod}\ N),\label{timevoPhase}
\end{equation}
where $\mathcal{O}(0)$ is some vector of the form \eqref{vecgen} representing the initial operator and $M$ is a $2 \times 2$ matrix representing the rule. The $\omega^\varphi$ factor is an overall phase coming from reordering the factors as $\bigotimes_\alpha Q_\alpha^{i_\alpha} P_\alpha^{i_\alpha}$ after applying $M$.

Implementing a particular rule requires the associated matrix. The rule \eqref{ourRule} used in the main text maps to
\begin{equation}
M = \begin{pmatrix}
q^{-1} + 1 + q & N-1\\
1 & 0
\end{pmatrix}.\label{ourRuleMatrix}
\end{equation}

\subsection{Squared Commutator in Clifford QCAs}\label{app:dcOscillatory}

We now derive \eqref{doubleCommCQCA}. Clifford QCAs are invariant under translations. We also assume invariance under reflections, making the rule \textit{palindromic} as in \cite{Gutschow_2009,Gutschow_2010}. Then, we may shift the initial operators \eqref{initOps} by $-\alpha$ and reflect them around $0$ to write the squared commutator at $\alpha$ as
\begin{equation}
C_\alpha(t) = \expval{[W_0(t),V_\alpha]^\dagger[W_0(t),V_\alpha]},
\end{equation}
where $V_\alpha = \cdots \otimes \mathds{1}_{\alpha-1} \otimes \Sigma_\alpha \otimes \mathds{1}_{\alpha+1} \otimes \cdots$. This is the expression computed in our simulation.

All generalized Pauli matrices may be written as products of $Q$'s and $P$'s. Additionally, in terms of the Laurent polynomial notation \eqref{vecgen}, we write the entries of the $t$-fold product of the rule matrix $M$ in index notation as
\begin{equation}
\left(M^t\right)_{IJ} = \sum_{\alpha = -\infty}^\infty \xi_{IJ,\alpha}(t) q^\alpha,\label{genMat}
\end{equation}
where the indices $I,J$ run over $Q$ and $P$ and each $\xi_{IJ,\alpha}(t)$ is an integer mod $N$. Meanwhile, supposing $\widetilde{\Sigma}_0 = Q^{i} P^{j}$, we write $W_0$ in index notation as
\begin{equation}
\left(W_0\right)_I = i \delta_{I,Q} + j \delta_{I,P},
\end{equation}
And so, up to an overall phase the time-evolved operator $W_0(t)$ is
\begin{equation}
\begin{split}
\left(M^t W_0\right)_{I}
&= \sum_{J} \left[\sum_{\alpha} \xi_{IJ,\alpha} (t) q^\alpha\right]\left(i \delta_{J,Q} + j \delta_{J,P}\right)\\
&= \sum_{\alpha} \left[i\xi_{IQ,\alpha}(t) + j\xi_{IP,\alpha}(t)\right]q^\alpha.
\end{split}\label{timeevoindex}
\end{equation}
We are ready to compute $[W_0(t),V_\alpha]$. Generalized Pauli matrices supported on different lattice sites commute, so we only need the factor of $M^t W_0$ at the $\alpha$ lattice site, defined as $\widetilde{\Sigma}_\alpha(t)$. From \eqref{timeevoindex}, this is
\begin{equation}
\begin{split}
\widetilde{\Sigma}_\alpha(t) &= Q^{i\xi_{QQ,\alpha}(t) + j\xi_{QP,\alpha}(t)} P^{i\xi_{PQ,\alpha}(t) + j\xi_{PP,\alpha}(t)}\\
&\equiv Q^{A(t)} P^{B(t)}.
\end{split}
\end{equation}
For brevity, we have redefined the exponents as integral functions $A(t)$ and $B(t)$. Additionally, we write the factor $\Sigma_\alpha$ of $V_\alpha$ as $Q^C P^D$. Note that $A,B,C,D$ also have $\alpha$ dependence, which we leave implicit for now. From the Clifford algebra \eqref{clifford}, the commutator is
\begin{align}
[W_0(t),V_\alpha]
&= \omega^{\varphi(t)} [\widetilde{\Sigma}_\alpha(t),\Sigma_\alpha]\\
&= \omega^{\varphi(t)} \left[\omega^{B(t) C} - \omega^{A(t) D}\right]Q^{A(t) + C} P^{B(t) + D}.\nonumber
\end{align}
Recall that $\omega = e^{2\pi i/N}$, and so $\omega^* = \omega^{-1}$. Additionally, the generalized Pauli matrices are unitary. So, we have
\begin{equation}
[W_0(t),V_\alpha]^\dagger [W_0(t),V_\alpha] = 4\sin^2\left[\frac{\pi}{N}\xi(\alpha,t)\right]\mathds{1},\label{sqcommMicro}
\end{equation}
where we have defined $\xi(\alpha,t) = A(t) D - B(t) C$ (making $\alpha$ explicit again) and exploited various trigonometric identities. The overall phase $\omega^{\varphi(t)}$ from \eqref{timevoPhase} cancels. The expectation value of \eqref{sqcommMicro} is \eqref{doubleCommCQCA}.

We only assume a palindromic rule to derive the sinusoidal expression. In the main text, we specify \eqref{ourRuleMatrix} and find jumps in the scrambling time at specific values of $N$ corresponding to a particular sequence of Whitney numbers \eqref{whitneyOurs}. However, such jumps are a generic feature of \eqref{doubleCommCQCA}, with the corresponding values of $N$ related to the sequence $\left(\xi(0,t)\right)_{t \geq 0}$. It would be interesting to explore analogous sequences from other palindromic rules.

\subsection{Fractal Dimension and Box Counting}

The dimension of a fractal can be found through a method called \textit{box counting}. The general procedure is as follows. Suppose that we have some shape of area $\mathcal{V}$. We then define a unit length $\varepsilon$ and cover the shape with $\mathcal{N}$ ``boxes" of area $\varepsilon^{D}$. There exists some $D$ for which
\begin{equation}
\lim_{\varepsilon \to 0}\mathcal{N} \varepsilon^D = \mathcal{V}.
\end{equation}
Essentially, we want to find this $D$. To do so, note that in the $\varepsilon \to 0$ limit we may write
\begin{equation}
\log \mathcal{N} \sim D \log \left(\frac{1}{\varepsilon}\right) + \log\mathcal{V},
\end{equation}
and so $D$ is the slope of the $\log\mathcal{N}$ vs. $\log(1/\varepsilon)$ line near $\varepsilon  = 0$. The $\log\mathcal{V}$ term can be ignored.

\begin{figure}
\centering
\includegraphics[scale=0.25]{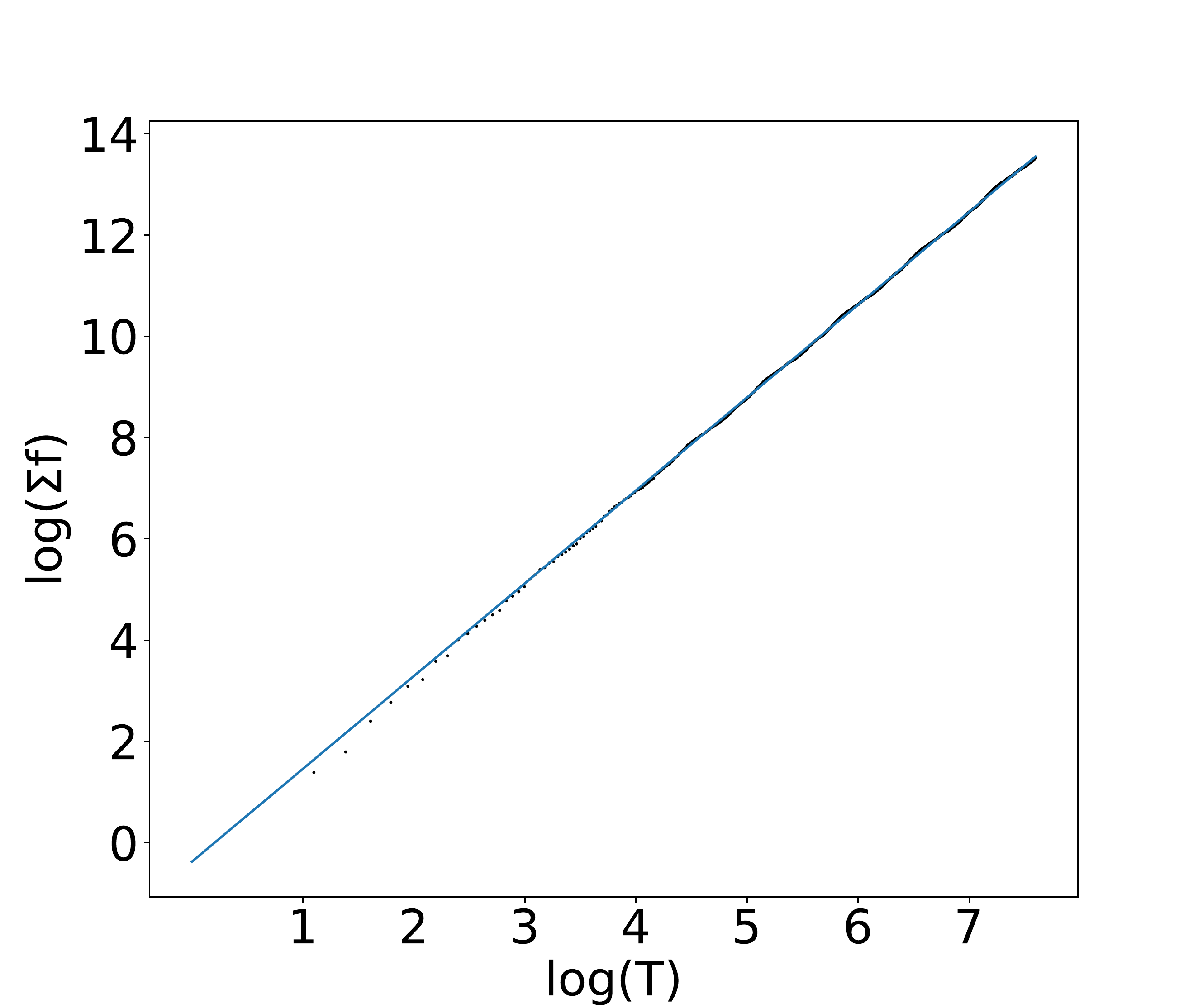}
\caption{The plot of $\log\left(\sum f\right)$ vs. $\log T$ for the Clifford QCA with initialization $\Sigma_0 = Q_0$, $\widetilde{\Sigma}_{\alpha} = Q_\alpha$ and $N = 2$. This ``box-counting plot" can be used to obtain the fractal dimension of the space-time plot of the squared commutator. The dimension is identified as the slope of the line at large $T$, which we find to be $\sim 1.83$.}
\label{figs:fracDim}
\end{figure}

In our QCAs, we implement an adaptation of box counting. Define $f(t)$ as the number of lattice sites at $t$ for which the squared commutator is $O(1)$. We simulate the QCA up to time $t = T$. We associate
\begin{equation}
\mathcal{N} = \sum_{t = 0}^{T} f(t),\ \ \varepsilon = \frac{1}{T}.
\end{equation}
And so, by plotting $\log \left(\sum f\right)$ vs. $\log T$, we can identify the large-scale structure's fractal dimension $D$ as the large-$T$ slope. See for example Figure \ref{figs:fracDim}, which shows the plot associated with Figure \ref{figs:spacetimeN2} (top left).

We also emphasize that this is not the only way to compute the fractal dimension. 

\bibliographystyle{apsrev4-2}
\bibliography{refs.bib}

\begin{thebibliography}{60}%
\makeatletter
\providecommand \@ifxundefined [1]{%
 \@ifx{#1\undefined}
}%
\providecommand \@ifnum [1]{%
 \ifnum #1\expandafter \@firstoftwo
 \else \expandafter \@secondoftwo
 \fi
}%
\providecommand \@ifx [1]{%
 \ifx #1\expandafter \@firstoftwo
 \else \expandafter \@secondoftwo
 \fi
}%
\providecommand \natexlab [1]{#1}%
\providecommand \enquote  [1]{``#1''}%
\providecommand \bibnamefont  [1]{#1}%
\providecommand \bibfnamefont [1]{#1}%
\providecommand \citenamefont [1]{#1}%
\providecommand \href@noop [0]{\@secondoftwo}%
\providecommand \href [0]{\begingroup \@sanitize@url \@href}%
\providecommand \@href[1]{\@@startlink{#1}\@@href}%
\providecommand \@@href[1]{\endgroup#1\@@endlink}%
\providecommand \@sanitize@url [0]{\catcode `\\12\catcode `\$12\catcode
  `\&12\catcode `\#12\catcode `\^12\catcode `\_12\catcode `\%12\relax}%
\providecommand \@@startlink[1]{}%
\providecommand \@@endlink[0]{}%
\providecommand \url  [0]{\begingroup\@sanitize@url \@url }%
\providecommand \@url [1]{\endgroup\@href {#1}{\urlprefix }}%
\providecommand \urlprefix  [0]{URL }%
\providecommand \Eprint [0]{\href }%
\providecommand \doibase [0]{https://doi.org/}%
\providecommand \selectlanguage [0]{\@gobble}%
\providecommand \bibinfo  [0]{\@secondoftwo}%
\providecommand \bibfield  [0]{\@secondoftwo}%
\providecommand \translation [1]{[#1]}%
\providecommand \BibitemOpen [0]{}%
\providecommand \bibitemStop [0]{}%
\providecommand \bibitemNoStop [0]{.\EOS\space}%
\providecommand \EOS [0]{\spacefactor3000\relax}%
\providecommand \BibitemShut  [1]{\csname bibitem#1\endcsname}%
\let\auto@bib@innerbib\@empty
\bibitem [{\citenamefont {Hayden}\ and\ \citenamefont
  {Preskill}(2007)}]{Hayden:2007cs}%
  \BibitemOpen
  \bibfield  {author} {\bibinfo {author} {\bibfnamefont {P.}~\bibnamefont
  {Hayden}}\ and\ \bibinfo {author} {\bibfnamefont {J.}~\bibnamefont
  {Preskill}},\ }\href {https://doi.org/10.1088/1126-6708/2007/09/120}
  {\bibfield  {journal} {\bibinfo  {journal} {JHEP}\ }\textbf {\bibinfo
  {volume} {09}}\bibfield  {number} {\bibinfo  {number} { (2007)},\ \bibinfo
  {pages} {120}},\ }\Eprint {https://arxiv.org/abs/0708.4025} {arXiv:0708.4025
  [hep-th]} \BibitemShut {NoStop}%
\bibitem [{\citenamefont {Sekino}\ and\ \citenamefont
  {Susskind}(2008)}]{Sekino:2008he}%
  \BibitemOpen
  \bibfield  {author} {\bibinfo {author} {\bibfnamefont {Y.}~\bibnamefont
  {Sekino}}\ and\ \bibinfo {author} {\bibfnamefont {L.}~\bibnamefont
  {Susskind}},\ }\href {https://doi.org/10.1088/1126-6708/2008/10/065}
  {\bibfield  {journal} {\bibinfo  {journal} {JHEP}\ }\textbf {\bibinfo
  {volume} {10}}\bibfield  {number} {\bibinfo  {number} { (2008)},\ \bibinfo
  {pages} {065}},\ }\Eprint {https://arxiv.org/abs/0808.2096} {arXiv:0808.2096
  [hep-th]} \BibitemShut {NoStop}%
\bibitem [{\citenamefont {Shenker}\ and\ \citenamefont
  {Stanford}(2014)}]{Shenker:2013pqa}%
  \BibitemOpen
  \bibfield  {author} {\bibinfo {author} {\bibfnamefont {S.~H.}\ \bibnamefont
  {Shenker}}\ and\ \bibinfo {author} {\bibfnamefont {D.}~\bibnamefont
  {Stanford}},\ }\href {https://doi.org/10.1007/JHEP03(2014)067} {\bibfield
  {journal} {\bibinfo  {journal} {JHEP}\ }\textbf {\bibinfo {volume}
  {03}}\bibfield  {number} {\bibinfo  {number} { (2014)},\ \bibinfo {pages}
  {067}},\ }\Eprint {https://arxiv.org/abs/1306.0622} {arXiv:1306.0622
  [hep-th]} \BibitemShut {NoStop}%
\bibitem [{\citenamefont {Deutsch}(1991)}]{Deutsch:1991eth}%
  \BibitemOpen
  \bibfield  {author} {\bibinfo {author} {\bibfnamefont {J.~M.}\ \bibnamefont
  {Deutsch}},\ }\href {https://doi.org/10.1103/PhysRevA.43.2046} {\bibfield
  {journal} {\bibinfo  {journal} {Phys. Rev. A}\ }\textbf {\bibinfo {volume}
  {43}},\ \bibinfo {pages} {2046} (\bibinfo {year} {1991})}\BibitemShut
  {NoStop}%
\bibitem [{\citenamefont {Srednicki}(1994)}]{Srednicki:1994mfb}%
  \BibitemOpen
  \bibfield  {author} {\bibinfo {author} {\bibfnamefont {M.}~\bibnamefont
  {Srednicki}},\ }\href {https://doi.org/10.1103/physreve.50.888} {\bibfield
  {journal} {\bibinfo  {journal} {Phys. Rev. E}\ }\textbf {\bibinfo {volume}
  {50}},\ \bibinfo {pages} {888} (\bibinfo {year} {1994})}\BibitemShut
  {NoStop}%
\bibitem [{\citenamefont {{Tasaki}}(1998)}]{Tasaki:1997ran}%
  \BibitemOpen
  \bibfield  {author} {\bibinfo {author} {\bibfnamefont {H.}~\bibnamefont
  {{Tasaki}}},\ }\href {https://doi.org/10.1103/PhysRevLett.80.1373} {\bibfield
   {journal} {\bibinfo  {journal} {Phys. Rev. Lett.}\ }\textbf {\bibinfo
  {volume} {80}},\ \bibinfo {pages} {1373} (\bibinfo {year} {1998})},\ \Eprint
  {https://arxiv.org/abs/cond-mat/9707253} {arXiv:cond-mat/9707253}
  \BibitemShut {NoStop}%
\bibitem [{\citenamefont {{Rigol}}\ \emph {et~al.}(2008)\citenamefont
  {{Rigol}}, \citenamefont {{Dunjko}},\ and\ \citenamefont
  {{Olshanii}}}]{Rigol:2007thm}%
  \BibitemOpen
  \bibfield  {author} {\bibinfo {author} {\bibfnamefont {M.}~\bibnamefont
  {{Rigol}}}, \bibinfo {author} {\bibfnamefont {V.}~\bibnamefont {{Dunjko}}},\
  and\ \bibinfo {author} {\bibfnamefont {M.}~\bibnamefont {{Olshanii}}},\
  }\href {https://doi.org/10.1038/nature06838} {\bibfield  {journal} {\bibinfo
  {journal} {Nature}\ }\textbf {\bibinfo {volume} {452}},\ \bibinfo {pages}
  {854} (\bibinfo {year} {2008})},\ \Eprint {https://arxiv.org/abs/0708.1324}
  {arXiv:0708.1324 [cond-mat.stat-mech]} \BibitemShut {NoStop}%
\bibitem [{\citenamefont {Heller}(1984)}]{Heller:1984zz}%
  \BibitemOpen
  \bibfield  {author} {\bibinfo {author} {\bibfnamefont {E.~J.}\ \bibnamefont
  {Heller}},\ }\href {https://doi.org/10.1103/PhysRevLett.53.1515} {\bibfield
  {journal} {\bibinfo  {journal} {Phys. Rev. Lett.}\ }\textbf {\bibinfo
  {volume} {53}},\ \bibinfo {pages} {1515} (\bibinfo {year}
  {1984})}\BibitemShut {NoStop}%
\bibitem [{\citenamefont {Turner}\ \emph {et~al.}(2018)\citenamefont {Turner},
  \citenamefont {Michailidis}, \citenamefont {Abanin}, \citenamefont {Serbyn},\
  and\ \citenamefont {Papi{\'c}}}]{Turner2017WeakEB}%
  \BibitemOpen
  \bibfield  {author} {\bibinfo {author} {\bibfnamefont {C.~J.}\ \bibnamefont
  {Turner}}, \bibinfo {author} {\bibfnamefont {A.~A.}\ \bibnamefont
  {Michailidis}}, \bibinfo {author} {\bibfnamefont {D.~A.}\ \bibnamefont
  {Abanin}}, \bibinfo {author} {\bibfnamefont {M.}~\bibnamefont {Serbyn}},\
  and\ \bibinfo {author} {\bibfnamefont {Z.}~\bibnamefont {Papi{\'c}}},\ }\href
  {https://doi.org/10.1038/s41567-018-0137-5} {\bibfield  {journal} {\bibinfo
  {journal} {Nature Physics}\ }\textbf {\bibinfo {volume} {14}},\ \bibinfo
  {pages} {745} (\bibinfo {year} {2018})}\BibitemShut {NoStop}%
\bibitem [{\citenamefont {Maldacena}\ \emph {et~al.}(2016)\citenamefont
  {Maldacena}, \citenamefont {Shenker},\ and\ \citenamefont
  {Stanford}}]{Maldacena:2015waa}%
  \BibitemOpen
  \bibfield  {author} {\bibinfo {author} {\bibfnamefont {J.}~\bibnamefont
  {Maldacena}}, \bibinfo {author} {\bibfnamefont {S.~H.}\ \bibnamefont
  {Shenker}},\ and\ \bibinfo {author} {\bibfnamefont {D.}~\bibnamefont
  {Stanford}},\ }\href {https://doi.org/10.1007/JHEP08(2016)106} {\bibfield
  {journal} {\bibinfo  {journal} {JHEP}\ }\textbf {\bibinfo {volume}
  {08}}\bibfield  {number} {\bibinfo  {number} { (2016)},\ \bibinfo {pages}
  {106}},\ }\Eprint {https://arxiv.org/abs/1503.01409} {arXiv:1503.01409
  [hep-th]} \BibitemShut {NoStop}%
\bibitem [{\citenamefont {Faulkner}\ \emph {et~al.}(2022)\citenamefont
  {Faulkner}, \citenamefont {Hartman}, \citenamefont {Headrick}, \citenamefont
  {Rangamani},\ and\ \citenamefont {Swingle}}]{Faulkner:2022mlp}%
  \BibitemOpen
  \bibfield  {author} {\bibinfo {author} {\bibfnamefont {T.}~\bibnamefont
  {Faulkner}}, \bibinfo {author} {\bibfnamefont {T.}~\bibnamefont {Hartman}},
  \bibinfo {author} {\bibfnamefont {M.}~\bibnamefont {Headrick}}, \bibinfo
  {author} {\bibfnamefont {M.}~\bibnamefont {Rangamani}},\ and\ \bibinfo
  {author} {\bibfnamefont {B.}~\bibnamefont {Swingle}},\ }in\ \href@noop {}
  {\emph {\bibinfo {booktitle} {{2022 Snowmass Summer Study}}}}\ (\bibinfo
  {year} {2022})\ \Eprint {https://arxiv.org/abs/2203.07117} {arXiv:2203.07117
  [hep-th]} \BibitemShut {NoStop}%
\bibitem [{\citenamefont {Susskind}(2016)}]{Susskind:2014moa}%
  \BibitemOpen
  \bibfield  {author} {\bibinfo {author} {\bibfnamefont {L.}~\bibnamefont
  {Susskind}},\ }\href {https://doi.org/10.1002/prop.201500095} {\bibfield
  {journal} {\bibinfo  {journal} {Fortsch. Phys.}\ }\textbf {\bibinfo {volume}
  {64}},\ \bibinfo {pages} {49} (\bibinfo {year} {2016})},\ \Eprint
  {https://arxiv.org/abs/1411.0690} {arXiv:1411.0690 [hep-th]} \BibitemShut
  {NoStop}%
\bibitem [{\citenamefont {Stanford}\ and\ \citenamefont
  {Susskind}(2014)}]{Stanford:2014jda}%
  \BibitemOpen
  \bibfield  {author} {\bibinfo {author} {\bibfnamefont {D.}~\bibnamefont
  {Stanford}}\ and\ \bibinfo {author} {\bibfnamefont {L.}~\bibnamefont
  {Susskind}},\ }\href {https://doi.org/10.1103/PhysRevD.90.126007} {\bibfield
  {journal} {\bibinfo  {journal} {Phys. Rev. D}\ }\textbf {\bibinfo {volume}
  {90}},\ \bibinfo {pages} {126007} (\bibinfo {year} {2014})},\ \Eprint
  {https://arxiv.org/abs/1406.2678} {arXiv:1406.2678 [hep-th]} \BibitemShut
  {NoStop}%
\bibitem [{\citenamefont {Belin}\ \emph {et~al.}(2022)\citenamefont {Belin},
  \citenamefont {Myers}, \citenamefont {Ruan}, \citenamefont {S\'arosi},\ and\
  \citenamefont {Speranza}}]{Belin:2021bga}%
  \BibitemOpen
  \bibfield  {author} {\bibinfo {author} {\bibfnamefont {A.}~\bibnamefont
  {Belin}}, \bibinfo {author} {\bibfnamefont {R.~C.}\ \bibnamefont {Myers}},
  \bibinfo {author} {\bibfnamefont {S.-M.}\ \bibnamefont {Ruan}}, \bibinfo
  {author} {\bibfnamefont {G.}~\bibnamefont {S\'arosi}},\ and\ \bibinfo
  {author} {\bibfnamefont {A.~J.}\ \bibnamefont {Speranza}},\ }\href
  {https://doi.org/10.1103/PhysRevLett.128.081602} {\bibfield  {journal}
  {\bibinfo  {journal} {Phys. Rev. Lett.}\ }\textbf {\bibinfo {volume} {128}},\
  \bibinfo {pages} {081602} (\bibinfo {year} {2022})},\ \Eprint
  {https://arxiv.org/abs/2111.02429} {arXiv:2111.02429 [hep-th]} \BibitemShut
  {NoStop}%
\bibitem [{\citenamefont {Hosur}\ \emph {et~al.}(2016)\citenamefont {Hosur},
  \citenamefont {Qi}, \citenamefont {Roberts},\ and\ \citenamefont
  {Yoshida}}]{Hosur:2015ylk}%
  \BibitemOpen
  \bibfield  {author} {\bibinfo {author} {\bibfnamefont {P.}~\bibnamefont
  {Hosur}}, \bibinfo {author} {\bibfnamefont {X.-L.}\ \bibnamefont {Qi}},
  \bibinfo {author} {\bibfnamefont {D.~A.}\ \bibnamefont {Roberts}},\ and\
  \bibinfo {author} {\bibfnamefont {B.}~\bibnamefont {Yoshida}},\ }\href
  {https://doi.org/10.1007/JHEP02(2016)004} {\bibfield  {journal} {\bibinfo
  {journal} {JHEP}\ }\textbf {\bibinfo {volume} {02}}\bibfield  {number}
  {\bibinfo  {number} { (2016)},\ \bibinfo {pages} {004}},\ }\Eprint
  {https://arxiv.org/abs/1511.04021} {arXiv:1511.04021 [hep-th]} \BibitemShut
  {NoStop}%
\bibitem [{\citenamefont {Nahum}\ \emph {et~al.}(2018)\citenamefont {Nahum},
  \citenamefont {Vijay},\ and\ \citenamefont {Haah}}]{Nahum:2017yvy}%
  \BibitemOpen
  \bibfield  {author} {\bibinfo {author} {\bibfnamefont {A.}~\bibnamefont
  {Nahum}}, \bibinfo {author} {\bibfnamefont {S.}~\bibnamefont {Vijay}},\ and\
  \bibinfo {author} {\bibfnamefont {J.}~\bibnamefont {Haah}},\ }\href
  {https://doi.org/10.1103/PhysRevX.8.021014} {\bibfield  {journal} {\bibinfo
  {journal} {Phys. Rev. X}\ }\textbf {\bibinfo {volume} {8}},\ \bibinfo {pages}
  {021014} (\bibinfo {year} {2018})},\ \Eprint
  {https://arxiv.org/abs/1705.08975} {arXiv:1705.08975 [cond-mat.str-el]}
  \BibitemShut {NoStop}%
\bibitem [{\citenamefont {Blake}(2016)}]{Blake:2016wvh}%
  \BibitemOpen
  \bibfield  {author} {\bibinfo {author} {\bibfnamefont {M.}~\bibnamefont
  {Blake}},\ }\href {https://doi.org/10.1103/PhysRevLett.117.091601} {\bibfield
   {journal} {\bibinfo  {journal} {Phys. Rev. Lett.}\ }\textbf {\bibinfo
  {volume} {117}},\ \bibinfo {pages} {091601} (\bibinfo {year} {2016})},\
  \Eprint {https://arxiv.org/abs/1603.08510} {arXiv:1603.08510 [hep-th]}
  \BibitemShut {NoStop}%
\bibitem [{\citenamefont {Davison}\ \emph {et~al.}(2017)\citenamefont
  {Davison}, \citenamefont {Fu}, \citenamefont {Georges}, \citenamefont {Gu},
  \citenamefont {Jensen},\ and\ \citenamefont {Sachdev}}]{Davison:2016ngz}%
  \BibitemOpen
  \bibfield  {author} {\bibinfo {author} {\bibfnamefont {R.~A.}\ \bibnamefont
  {Davison}}, \bibinfo {author} {\bibfnamefont {W.}~\bibnamefont {Fu}},
  \bibinfo {author} {\bibfnamefont {A.}~\bibnamefont {Georges}}, \bibinfo
  {author} {\bibfnamefont {Y.}~\bibnamefont {Gu}}, \bibinfo {author}
  {\bibfnamefont {K.}~\bibnamefont {Jensen}},\ and\ \bibinfo {author}
  {\bibfnamefont {S.}~\bibnamefont {Sachdev}},\ }\href
  {https://doi.org/10.1103/PhysRevB.95.155131} {\bibfield  {journal} {\bibinfo
  {journal} {Phys. Rev. B}\ }\textbf {\bibinfo {volume} {95}},\ \bibinfo
  {pages} {155131} (\bibinfo {year} {2017})},\ \Eprint
  {https://arxiv.org/abs/1612.00849} {arXiv:1612.00849 [cond-mat.str-el]}
  \BibitemShut {NoStop}%
\bibitem [{\citenamefont {Hartman}\ \emph {et~al.}(2017)\citenamefont
  {Hartman}, \citenamefont {Hartnoll},\ and\ \citenamefont
  {Mahajan}}]{Hartman:2017hhp}%
  \BibitemOpen
  \bibfield  {author} {\bibinfo {author} {\bibfnamefont {T.}~\bibnamefont
  {Hartman}}, \bibinfo {author} {\bibfnamefont {S.~A.}\ \bibnamefont
  {Hartnoll}},\ and\ \bibinfo {author} {\bibfnamefont {R.}~\bibnamefont
  {Mahajan}},\ }\href {https://doi.org/10.1103/PhysRevLett.119.141601}
  {\bibfield  {journal} {\bibinfo  {journal} {Phys. Rev. Lett.}\ }\textbf
  {\bibinfo {volume} {119}},\ \bibinfo {pages} {141601} (\bibinfo {year}
  {2017})},\ \Eprint {https://arxiv.org/abs/1706.00019} {arXiv:1706.00019
  [hep-th]} \BibitemShut {NoStop}%
\bibitem [{\citenamefont {Susskind}(2011)}]{Susskind:2011ap}%
  \BibitemOpen
  \bibfield  {author} {\bibinfo {author} {\bibfnamefont {L.}~\bibnamefont
  {Susskind}},\ }\href@noop {} {\  (\bibinfo {year} {2011})},\ \Eprint
  {https://arxiv.org/abs/1101.6048} {arXiv:1101.6048 [hep-th]} \BibitemShut
  {NoStop}%
\bibitem [{\citenamefont {Lashkari}\ \emph {et~al.}(2013)\citenamefont
  {Lashkari}, \citenamefont {Stanford}, \citenamefont {Hastings}, \citenamefont
  {Osborne},\ and\ \citenamefont {Hayden}}]{Lashkari:2011yi}%
  \BibitemOpen
  \bibfield  {author} {\bibinfo {author} {\bibfnamefont {N.}~\bibnamefont
  {Lashkari}}, \bibinfo {author} {\bibfnamefont {D.}~\bibnamefont {Stanford}},
  \bibinfo {author} {\bibfnamefont {M.}~\bibnamefont {Hastings}}, \bibinfo
  {author} {\bibfnamefont {T.}~\bibnamefont {Osborne}},\ and\ \bibinfo {author}
  {\bibfnamefont {P.}~\bibnamefont {Hayden}},\ }\href
  {https://doi.org/10.1007/JHEP04(2013)022} {\bibfield  {journal} {\bibinfo
  {journal} {JHEP}\ }\textbf {\bibinfo {volume} {04}}\bibfield  {number}
  {\bibinfo  {number} { (2013)},\ \bibinfo {pages} {022}},\ }\Eprint
  {https://arxiv.org/abs/1111.6580} {arXiv:1111.6580 [hep-th]} \BibitemShut
  {NoStop}%
\bibitem [{\citenamefont {Swingle}\ \emph {et~al.}(2016)\citenamefont
  {Swingle}, \citenamefont {Bentsen}, \citenamefont {Schleier-Smith},\ and\
  \citenamefont {Hayden}}]{Swingle:2016var}%
  \BibitemOpen
  \bibfield  {author} {\bibinfo {author} {\bibfnamefont {B.}~\bibnamefont
  {Swingle}}, \bibinfo {author} {\bibfnamefont {G.}~\bibnamefont {Bentsen}},
  \bibinfo {author} {\bibfnamefont {M.}~\bibnamefont {Schleier-Smith}},\ and\
  \bibinfo {author} {\bibfnamefont {P.}~\bibnamefont {Hayden}},\ }\href
  {https://doi.org/10.1103/PhysRevA.94.040302} {\bibfield  {journal} {\bibinfo
  {journal} {Phys. Rev. A}\ }\textbf {\bibinfo {volume} {94}},\ \bibinfo
  {pages} {040302} (\bibinfo {year} {2016})},\ \Eprint
  {https://arxiv.org/abs/1602.06271} {arXiv:1602.06271 [quant-ph]} \BibitemShut
  {NoStop}%
\bibitem [{\citenamefont {Xu}\ and\ \citenamefont
  {Swingle}(2020)}]{Xu:2018xfz}%
  \BibitemOpen
  \bibfield  {author} {\bibinfo {author} {\bibfnamefont {S.}~\bibnamefont
  {Xu}}\ and\ \bibinfo {author} {\bibfnamefont {B.}~\bibnamefont {Swingle}},\
  }\href {https://doi.org/10.1038/s41567-019-0712-4} {\bibfield  {journal}
  {\bibinfo  {journal} {Nature Phys.}\ }\textbf {\bibinfo {volume} {16}},\
  \bibinfo {pages} {199} (\bibinfo {year} {2020})},\ \Eprint
  {https://arxiv.org/abs/1802.00801} {arXiv:1802.00801 [quant-ph]} \BibitemShut
  {NoStop}%
\bibitem [{\citenamefont {Xu}\ and\ \citenamefont
  {Swingle}(2022)}]{Xu:2022vko}%
  \BibitemOpen
  \bibfield  {author} {\bibinfo {author} {\bibfnamefont {S.}~\bibnamefont
  {Xu}}\ and\ \bibinfo {author} {\bibfnamefont {B.}~\bibnamefont {Swingle}},\
  }\href@noop {} {\  (\bibinfo {year} {2022})},\ \Eprint
  {https://arxiv.org/abs/2202.07060} {arXiv:2202.07060 [quant-ph]} \BibitemShut
  {NoStop}%
\bibitem [{\citenamefont {Li}\ \emph {et~al.}(2017)\citenamefont {Li},
  \citenamefont {Fan}, \citenamefont {Wang}, \citenamefont {Ye}, \citenamefont
  {Zeng}, \citenamefont {Zhai}, \citenamefont {Peng},\ and\ \citenamefont
  {Du}}]{Li:2016xhw}%
  \BibitemOpen
  \bibfield  {author} {\bibinfo {author} {\bibfnamefont {J.}~\bibnamefont
  {Li}}, \bibinfo {author} {\bibfnamefont {R.}~\bibnamefont {Fan}}, \bibinfo
  {author} {\bibfnamefont {H.}~\bibnamefont {Wang}}, \bibinfo {author}
  {\bibfnamefont {B.}~\bibnamefont {Ye}}, \bibinfo {author} {\bibfnamefont
  {B.}~\bibnamefont {Zeng}}, \bibinfo {author} {\bibfnamefont {H.}~\bibnamefont
  {Zhai}}, \bibinfo {author} {\bibfnamefont {X.}~\bibnamefont {Peng}},\ and\
  \bibinfo {author} {\bibfnamefont {J.}~\bibnamefont {Du}},\ }\href
  {https://doi.org/10.1103/PhysRevX.7.031011} {\bibfield  {journal} {\bibinfo
  {journal} {Phys. Rev. X}\ }\textbf {\bibinfo {volume} {7}},\ \bibinfo {pages}
  {031011} (\bibinfo {year} {2017})},\ \Eprint
  {https://arxiv.org/abs/1609.01246} {arXiv:1609.01246 [cond-mat.str-el]}
  \BibitemShut {NoStop}%
\bibitem [{\citenamefont {Zhu}\ \emph {et~al.}(2022)\citenamefont {Zhu} \emph
  {et~al.}}]{Zhu:2021uzs}%
  \BibitemOpen
  \bibfield  {author} {\bibinfo {author} {\bibfnamefont {Q.}~\bibnamefont
  {Zhu}} \emph {et~al.},\ }\href
  {https://doi.org/10.1103/PhysRevLett.128.160502} {\bibfield  {journal}
  {\bibinfo  {journal} {Phys. Rev. Lett.}\ }\textbf {\bibinfo {volume} {128}},\
  \bibinfo {pages} {160502} (\bibinfo {year} {2022})},\ \Eprint
  {https://arxiv.org/abs/2101.08031} {arXiv:2101.08031 [quant-ph]} \BibitemShut
  {NoStop}%
\bibitem [{\citenamefont {{Lent}}\ \emph {et~al.}(1993)\citenamefont {{Lent}},
  \citenamefont {{Tougaw}}, \citenamefont {{Porod}},\ and\ \citenamefont
  {{Bernstein}}}]{Lent:1993qca}%
  \BibitemOpen
  \bibfield  {author} {\bibinfo {author} {\bibfnamefont {C.~S.}\ \bibnamefont
  {{Lent}}}, \bibinfo {author} {\bibfnamefont {P.~D.}\ \bibnamefont
  {{Tougaw}}}, \bibinfo {author} {\bibfnamefont {W.}~\bibnamefont {{Porod}}},\
  and\ \bibinfo {author} {\bibfnamefont {G.~H.}\ \bibnamefont {{Bernstein}}},\
  }\href {https://doi.org/10.1088/0957-4484/4/1/004} {\bibfield  {journal}
  {\bibinfo  {journal} {Nanotechnology}\ }\textbf {\bibinfo {volume} {4}},\
  \bibinfo {pages} {49} (\bibinfo {year} {1993})}\BibitemShut {NoStop}%
\bibitem [{\citenamefont {Schumacher}\ and\ \citenamefont
  {Werner}(2004)}]{Schumacher:2004qca}%
  \BibitemOpen
  \bibfield  {author} {\bibinfo {author} {\bibfnamefont {B.}~\bibnamefont
  {Schumacher}}\ and\ \bibinfo {author} {\bibfnamefont {R.~F.}\ \bibnamefont
  {Werner}},\ }\href@noop {} {\  (\bibinfo {year} {2004})},\ \Eprint
  {https://arxiv.org/abs/quant-ph/0405174} {arXiv:quant-ph/0405174 [quant-ph]}
  \BibitemShut {NoStop}%
\bibitem [{\citenamefont {Farrelly}(2020)}]{Farrelly:2019zds}%
  \BibitemOpen
  \bibfield  {author} {\bibinfo {author} {\bibfnamefont {T.}~\bibnamefont
  {Farrelly}},\ }\href {https://doi.org/10.22331/q-2020-11-30-368} {\bibfield
  {journal} {\bibinfo  {journal} {Quantum}\ }\textbf {\bibinfo {volume} {4}},\
  \bibinfo {pages} {368} (\bibinfo {year} {2020})},\ \Eprint
  {https://arxiv.org/abs/1904.13318} {arXiv:1904.13318 [quant-ph]} \BibitemShut
  {NoStop}%
\bibitem [{\citenamefont {{Gopalakrishnan}}\ and\ \citenamefont
  {{Zakirov}}(2018)}]{Gopalkrishnan_2018}%
  \BibitemOpen
  \bibfield  {author} {\bibinfo {author} {\bibfnamefont {S.}~\bibnamefont
  {{Gopalakrishnan}}}\ and\ \bibinfo {author} {\bibfnamefont {B.}~\bibnamefont
  {{Zakirov}}},\ }\href
  {https://doi.org/10.1088/2058-9565/aad75910.48550/arXiv.1802.07729}
  {\bibfield  {journal} {\bibinfo  {journal} {Quantum Science and Technology}\
  }\textbf {\bibinfo {volume} {3}},\ \bibinfo {pages} {044004} (\bibinfo {year}
  {2018})},\ \Eprint {https://arxiv.org/abs/1802.07729} {arXiv:1802.07729
  [cond-mat.stat-mech]} \BibitemShut {NoStop}%
\bibitem [{\citenamefont {Gopalakrishnan}(2018)}]{Gopalakrishnan_20182}%
  \BibitemOpen
  \bibfield  {author} {\bibinfo {author} {\bibfnamefont {S.}~\bibnamefont
  {Gopalakrishnan}},\ }\bibfield  {journal} {\bibinfo  {journal} {Physical
  Review B}\ }\textbf {\bibinfo {volume} {98}},\ \href
  {https://doi.org/10.1103/PhysRevB.98.060302} {10.1103/PhysRevB.98.060302}
  (\bibinfo {year} {2018}),\ \Eprint {https://arxiv.org/abs/1806.04156}
  {arXiv:1806.04156 [cond-mat.stat-mech]} \BibitemShut {NoStop}%
\bibitem [{\citenamefont {Iadecola}\ and\ \citenamefont
  {Vijay}(2020)}]{Iadecola_2020}%
  \BibitemOpen
  \bibfield  {author} {\bibinfo {author} {\bibfnamefont {T.}~\bibnamefont
  {Iadecola}}\ and\ \bibinfo {author} {\bibfnamefont {S.}~\bibnamefont
  {Vijay}},\ }\bibfield  {journal} {\bibinfo  {journal} {Physical Review B}\
  }\textbf {\bibinfo {volume} {102}},\ \href
  {https://doi.org/10.1103/PhysRevB.102.180302} {10.1103/PhysRevB.102.180302}
  (\bibinfo {year} {2020}),\ \Eprint {https://arxiv.org/abs/2006.02440}
  {arXiv:2006.02440 [cond-mat.str-el]} \BibitemShut {NoStop}%
\bibitem [{\citenamefont {Hillberry}\ \emph {et~al.}(2021)\citenamefont
  {Hillberry}, \citenamefont {Jones}, \citenamefont {Vargas}, \citenamefont
  {Rall}, \citenamefont {Halpern}, \citenamefont {Bao}, \citenamefont
  {Notarnicola}, \citenamefont {Montangero},\ and\ \citenamefont
  {Carr}}]{Hillberry:2020nfj}%
  \BibitemOpen
  \bibfield  {author} {\bibinfo {author} {\bibfnamefont {L.~E.}\ \bibnamefont
  {Hillberry}}, \bibinfo {author} {\bibfnamefont {M.~T.}\ \bibnamefont
  {Jones}}, \bibinfo {author} {\bibfnamefont {D.~L.}\ \bibnamefont {Vargas}},
  \bibinfo {author} {\bibfnamefont {P.}~\bibnamefont {Rall}}, \bibinfo {author}
  {\bibfnamefont {N.~Y.}\ \bibnamefont {Halpern}}, \bibinfo {author}
  {\bibfnamefont {N.}~\bibnamefont {Bao}}, \bibinfo {author} {\bibfnamefont
  {S.}~\bibnamefont {Notarnicola}}, \bibinfo {author} {\bibfnamefont
  {S.}~\bibnamefont {Montangero}},\ and\ \bibinfo {author} {\bibfnamefont
  {L.~D.}\ \bibnamefont {Carr}},\ }\href
  {https://doi.org/10.1088/2058-9565/ac1c41} {\bibfield  {journal} {\bibinfo
  {journal} {Quantum Sci. Technol.}\ }\textbf {\bibinfo {volume} {6}},\
  \bibinfo {pages} {045017} (\bibinfo {year} {2021})},\ \Eprint
  {https://arxiv.org/abs/2005.01763} {arXiv:2005.01763 [quant-ph]} \BibitemShut
  {NoStop}%
\bibitem [{\citenamefont {Sellapillay}\ \emph {et~al.}(2022)\citenamefont
  {Sellapillay}, \citenamefont {Verga},\ and\ \citenamefont
  {Di~Molfetta}}]{Sellapillay:2022jkg}%
  \BibitemOpen
  \bibfield  {author} {\bibinfo {author} {\bibfnamefont {K.}~\bibnamefont
  {Sellapillay}}, \bibinfo {author} {\bibfnamefont {A.~D.}\ \bibnamefont
  {Verga}},\ and\ \bibinfo {author} {\bibfnamefont {G.}~\bibnamefont
  {Di~Molfetta}},\ }\href {https://doi.org/10.1103/PhysRevB.106.104309}
  {\bibfield  {journal} {\bibinfo  {journal} {Phys. Rev. B}\ }\textbf {\bibinfo
  {volume} {106}},\ \bibinfo {pages} {104309} (\bibinfo {year} {2022})},\
  \Eprint {https://arxiv.org/abs/2207.05360} {arXiv:2207.05360 [quant-ph]}
  \BibitemShut {NoStop}%
\bibitem [{\citenamefont {{Farshi}}\ \emph {et~al.}(2022)\citenamefont
  {{Farshi}}, \citenamefont {{Richter}}, \citenamefont {{Toniolo}},
  \citenamefont {{Pal}},\ and\ \citenamefont {{Masanes}}}]{Farshi:2022clf}%
  \BibitemOpen
  \bibfield  {author} {\bibinfo {author} {\bibfnamefont {T.}~\bibnamefont
  {{Farshi}}}, \bibinfo {author} {\bibfnamefont {J.}~\bibnamefont {{Richter}}},
  \bibinfo {author} {\bibfnamefont {D.}~\bibnamefont {{Toniolo}}}, \bibinfo
  {author} {\bibfnamefont {A.}~\bibnamefont {{Pal}}},\ and\ \bibinfo {author}
  {\bibfnamefont {L.}~\bibnamefont {{Masanes}}},\ }\href@noop {} {\  (\bibinfo
  {year} {2022})},\ \Eprint {https://arxiv.org/abs/2210.10129}
  {arXiv:2210.10129 [quant-ph]} \BibitemShut {NoStop}%
\bibitem [{\citenamefont {Sommers}\ \emph {et~al.}(2022)\citenamefont
  {Sommers}, \citenamefont {Huse},\ and\ \citenamefont
  {Gullans}}]{Sommers:2022cuz}%
  \BibitemOpen
  \bibfield  {author} {\bibinfo {author} {\bibfnamefont {G.~M.}\ \bibnamefont
  {Sommers}}, \bibinfo {author} {\bibfnamefont {D.~A.}\ \bibnamefont {Huse}},\
  and\ \bibinfo {author} {\bibfnamefont {M.~J.}\ \bibnamefont {Gullans}},\
  }\href@noop {} {\  (\bibinfo {year} {2022})},\ \Eprint
  {https://arxiv.org/abs/2210.10808} {arXiv:2210.10808 [quant-ph]} \BibitemShut
  {NoStop}%
\bibitem [{Note1()}]{Note1}%
  \BibitemOpen
  \bibinfo {note} {{\protect \textup {\hbox {\mathsurround \z@ \protect
  \normalfont (\ignorespaces \ref {boundComm}\unskip \@@italiccorr )}} and its
  bound in the holographic limit $p \to 0$ \cite {Roberts:2016wdl} are
  state-dependent, i.e. $\lambda _{\protect \text {L}}$ and $v_{\protect \text
  {B}}$ care about $V$ and $W(x)$. There is a similar, state-independent bound
  of this form with $p = 0$ called the \protect \textit {Lieb--Robinson bound}
  on the microscopic norm (as opposed to the average) of the squared commutator
  \cite {Lieb:1972wy,Nachtergaele_2006,Hastings_2006,Hastings:2010lrb}. \cite
  {Swingle:2016var} interprets the butterfly cone as a low-energy effective
  ``Lieb--Robinson" cone.}}\BibitemShut {Stop}%
\bibitem [{\citenamefont {Schlingemann}\ \emph {et~al.}(2008)\citenamefont
  {Schlingemann}, \citenamefont {Vogts},\ and\ \citenamefont
  {Werner}}]{Schlingemann_2008}%
  \BibitemOpen
  \bibfield  {author} {\bibinfo {author} {\bibfnamefont {D.-M.}\ \bibnamefont
  {Schlingemann}}, \bibinfo {author} {\bibfnamefont {H.}~\bibnamefont
  {Vogts}},\ and\ \bibinfo {author} {\bibfnamefont {R.~F.}\ \bibnamefont
  {Werner}},\ }\href {https://doi.org/10.1063/1.3005565} {\bibfield  {journal}
  {\bibinfo  {journal} {Journal of Mathematical Physics}\ }\textbf {\bibinfo
  {volume} {49}},\ \bibinfo {pages} {112104} (\bibinfo {year}
  {2008})}\BibitemShut {NoStop}%
\bibitem [{\citenamefont {Gütschow}(2009)}]{Gutschow_2009}%
  \BibitemOpen
  \bibfield  {author} {\bibinfo {author} {\bibfnamefont {J.}~\bibnamefont
  {Gütschow}},\ }\href {https://doi.org/10.1007/s00340-009-3840-1} {\bibfield
  {journal} {\bibinfo  {journal} {Applied Physics B}\ }\textbf {\bibinfo
  {volume} {98}},\ \bibinfo {pages} {623} (\bibinfo {year} {2009})}\BibitemShut
  {NoStop}%
\bibitem [{\citenamefont {Gütschow}\ \emph {et~al.}(2010)\citenamefont
  {Gütschow}, \citenamefont {Uphoff}, \citenamefont {Werner},\ and\
  \citenamefont {Zimbor{\'{a}}s}}]{Gutschow_2010}%
  \BibitemOpen
  \bibfield  {author} {\bibinfo {author} {\bibfnamefont {J.}~\bibnamefont
  {Gütschow}}, \bibinfo {author} {\bibfnamefont {S.}~\bibnamefont {Uphoff}},
  \bibinfo {author} {\bibfnamefont {R.~F.}\ \bibnamefont {Werner}},\ and\
  \bibinfo {author} {\bibfnamefont {Z.}~\bibnamefont {Zimbor{\'{a}}s}},\ }\href
  {https://doi.org/10.1063/1.3278513} {\bibfield  {journal} {\bibinfo
  {journal} {Journal of Mathematical Physics}\ }\textbf {\bibinfo {volume}
  {51}},\ \bibinfo {pages} {015203} (\bibinfo {year} {2010})}\BibitemShut
  {NoStop}%
\bibitem [{\citenamefont {Berenstein}(2018)}]{Berenstein:2018zif}%
  \BibitemOpen
  \bibfield  {author} {\bibinfo {author} {\bibfnamefont {D.}~\bibnamefont
  {Berenstein}},\ }\href@noop {} {\  (\bibinfo {year} {2018})},\ \Eprint
  {https://arxiv.org/abs/1803.02396} {arXiv:1803.02396 [hep-th]} \BibitemShut
  {NoStop}%
\bibitem [{\citenamefont {{de la Torre}}\ and\ \citenamefont
  {{Goyeneche}}(2003)}]{2003AmJPh..71...49D}%
  \BibitemOpen
  \bibfield  {author} {\bibinfo {author} {\bibfnamefont {A.~C.}\ \bibnamefont
  {{de la Torre}}}\ and\ \bibinfo {author} {\bibfnamefont {D.}~\bibnamefont
  {{Goyeneche}}},\ }\href {https://doi.org/10.1119/1.1514208} {\bibfield
  {journal} {\bibinfo  {journal} {American Journal of Physics}\ }\textbf
  {\bibinfo {volume} {71}},\ \bibinfo {pages} {49} (\bibinfo {year} {2003})},\
  \Eprint {https://arxiv.org/abs/quant-ph/0205159} {arXiv:quant-ph/0205159
  [quant-ph]} \BibitemShut {NoStop}%
\bibitem [{\citenamefont {G{\"u}tschow}\ \emph {et~al.}(2010)\citenamefont
  {G{\"u}tschow}, \citenamefont {Nesme},\ and\ \citenamefont
  {Werner}}]{Gtschow2010TheFS}%
  \BibitemOpen
  \bibfield  {author} {\bibinfo {author} {\bibfnamefont {J.}~\bibnamefont
  {G{\"u}tschow}}, \bibinfo {author} {\bibfnamefont {V.}~\bibnamefont
  {Nesme}},\ and\ \bibinfo {author} {\bibfnamefont {R.~F.}\ \bibnamefont
  {Werner}},\ }in\ \href@noop {} {\emph {\bibinfo {booktitle} {Automata}}}\
  (\bibinfo {year} {2010})\ \Eprint {https://arxiv.org/abs/1011.0313}
  {arXiv:1011.0313 [cs.DM]} \BibitemShut {NoStop}%
\bibitem [{\citenamefont {Berenstein}\ and\ \citenamefont
  {Kent}(2021)}]{Berenstein:2021lya}%
  \BibitemOpen
  \bibfield  {author} {\bibinfo {author} {\bibfnamefont {D.}~\bibnamefont
  {Berenstein}}\ and\ \bibinfo {author} {\bibfnamefont {B.}~\bibnamefont
  {Kent}},\ }\href@noop {} {\  (\bibinfo {year} {2021})},\ \Eprint
  {https://arxiv.org/abs/2107.12191} {arXiv:2107.12191 [cond-mat.stat-mech]}
  \BibitemShut {NoStop}%
\bibitem [{\citenamefont {Roberts}\ \emph {et~al.}(2015)\citenamefont
  {Roberts}, \citenamefont {Stanford},\ and\ \citenamefont
  {Susskind}}]{Roberts:2014isa}%
  \BibitemOpen
  \bibfield  {author} {\bibinfo {author} {\bibfnamefont {D.~A.}\ \bibnamefont
  {Roberts}}, \bibinfo {author} {\bibfnamefont {D.}~\bibnamefont {Stanford}},\
  and\ \bibinfo {author} {\bibfnamefont {L.}~\bibnamefont {Susskind}},\ }\href
  {https://doi.org/10.1007/JHEP03(2015)051} {\bibfield  {journal} {\bibinfo
  {journal} {JHEP}\ }\textbf {\bibinfo {volume} {03}}\bibfield  {number}
  {\bibinfo  {number} { (2015)},\ \bibinfo {pages} {051}},\ }\Eprint
  {https://arxiv.org/abs/1409.8180} {arXiv:1409.8180 [hep-th]} \BibitemShut
  {NoStop}%
\bibitem [{\citenamefont {Sloane}()}]{intseq}%
  \BibitemOpen
  \bibfield  {author} {\bibinfo {author} {\bibfnamefont {N.~J.~A.}\
  \bibnamefont {Sloane}},\ }\bibinfo {title} {A077419},\ in\ \href
  {https://oeis.org/} {\emph {\bibinfo {booktitle} {The On-Line Encyclopedia of
  Integer Sequences}}}\BibitemShut {NoStop}%
\bibitem [{\citenamefont {Munarini}\ and\ \citenamefont
  {Salvi}(2002)}]{Munarini2002OnTR}%
  \BibitemOpen
  \bibfield  {author} {\bibinfo {author} {\bibfnamefont {E.}~\bibnamefont
  {Munarini}}\ and\ \bibinfo {author} {\bibfnamefont {N.~Z.}\ \bibnamefont
  {Salvi}},\ }\href@noop {} {\bibfield  {journal} {\bibinfo  {journal}
  {Discret. Math.}\ }\textbf {\bibinfo {volume} {259}},\ \bibinfo {pages} {163}
  (\bibinfo {year} {2002})}\BibitemShut {NoStop}%
\bibitem [{\citenamefont {Conflitti}(2009)}]{Conflitti2005OnWN}%
  \BibitemOpen
  \bibfield  {author} {\bibinfo {author} {\bibfnamefont {A.}~\bibnamefont
  {Conflitti}},\ }\href {https://doi.org/10.1016/j.disc.2007.12.077} {\bibfield
   {journal} {\bibinfo  {journal} {Discret. Math.}\ }\textbf {\bibinfo {volume}
  {309}},\ \bibinfo {pages} {615} (\bibinfo {year} {2009})},\ \Eprint
  {https://arxiv.org/abs/math/0505636} {arXiv:math/0505636 [math]} \BibitemShut
  {NoStop}%
\bibitem [{Note2()}]{Note2}%
  \BibitemOpen
  \bibinfo {note} {{We can also define Whitney numbers for other partially
  ordered sets, such as ``crowns," which are cycilic fences \cite
  {Munarini2002OnTR,Conflitti2005OnWN}.}}\BibitemShut {Stop}%
\bibitem [{\citenamefont {Schwinger}(2001)}]{SchwingerBook}%
  \BibitemOpen
  \bibfield  {author} {\bibinfo {author} {\bibfnamefont {J.}~\bibnamefont
  {Schwinger}},\ }\href@noop {} {\emph {\bibinfo {title} {{Quantum Mechanics:
  Symbolism of Atomic Measurements}}}},\ edited by\ \bibinfo {editor}
  {\bibfnamefont {B.-G.}\ \bibnamefont {Englert}}\ (\bibinfo  {publisher}
  {Springer},\ \bibinfo {year} {2001})\BibitemShut {NoStop}%
\bibitem [{\citenamefont {Rabinovici}\ \emph {et~al.}(2021)\citenamefont
  {Rabinovici}, \citenamefont {S\'anchez-Garrido}, \citenamefont {Shir},\ and\
  \citenamefont {Sonner}}]{Rabinovici:2020ryf}%
  \BibitemOpen
  \bibfield  {author} {\bibinfo {author} {\bibfnamefont {E.}~\bibnamefont
  {Rabinovici}}, \bibinfo {author} {\bibfnamefont {A.}~\bibnamefont
  {S\'anchez-Garrido}}, \bibinfo {author} {\bibfnamefont {R.}~\bibnamefont
  {Shir}},\ and\ \bibinfo {author} {\bibfnamefont {J.}~\bibnamefont {Sonner}},\
  }\href {https://doi.org/10.1007/JHEP06(2021)062} {\bibfield  {journal}
  {\bibinfo  {journal} {JHEP}\ }\textbf {\bibinfo {volume} {06}}\bibfield
  {number} {\bibinfo  {number} { (2021)},\ \bibinfo {pages} {062}},\ }\Eprint
  {https://arxiv.org/abs/2009.01862} {arXiv:2009.01862 [hep-th]} \BibitemShut
  {NoStop}%
\bibitem [{\citenamefont {Bhattacharjee}\ \emph {et~al.}(2022)\citenamefont
  {Bhattacharjee}, \citenamefont {Sur},\ and\ \citenamefont
  {Nandy}}]{Bhattacharjee:2022qjw}%
  \BibitemOpen
  \bibfield  {author} {\bibinfo {author} {\bibfnamefont {B.}~\bibnamefont
  {Bhattacharjee}}, \bibinfo {author} {\bibfnamefont {S.}~\bibnamefont {Sur}},\
  and\ \bibinfo {author} {\bibfnamefont {P.}~\bibnamefont {Nandy}},\ }\href
  {https://doi.org/10.1103/PhysRevB.106.205150} {\bibfield  {journal} {\bibinfo
   {journal} {Phys. Rev. B}\ }\textbf {\bibinfo {volume} {106}},\ \bibinfo
  {pages} {205150} (\bibinfo {year} {2022})},\ \Eprint
  {https://arxiv.org/abs/2208.05503} {arXiv:2208.05503 [quant-ph]} \BibitemShut
  {NoStop}%
\bibitem [{\citenamefont {Liu}\ \emph {et~al.}(2021)\citenamefont {Liu},
  \citenamefont {Willsher}, \citenamefont {Bilitewski}, \citenamefont {Li},
  \citenamefont {Smith}, \citenamefont {Christensen}, \citenamefont
  {Moessner},\ and\ \citenamefont {Knolle}}]{Liu_2021}%
  \BibitemOpen
  \bibfield  {author} {\bibinfo {author} {\bibfnamefont {S.-W.}\ \bibnamefont
  {Liu}}, \bibinfo {author} {\bibfnamefont {J.}~\bibnamefont {Willsher}},
  \bibinfo {author} {\bibfnamefont {T.}~\bibnamefont {Bilitewski}}, \bibinfo
  {author} {\bibfnamefont {J.-J.}\ \bibnamefont {Li}}, \bibinfo {author}
  {\bibfnamefont {A.}~\bibnamefont {Smith}}, \bibinfo {author} {\bibfnamefont
  {K.}~\bibnamefont {Christensen}}, \bibinfo {author} {\bibfnamefont
  {R.}~\bibnamefont {Moessner}},\ and\ \bibinfo {author} {\bibfnamefont
  {J.}~\bibnamefont {Knolle}},\ }\bibfield  {journal} {\bibinfo  {journal}
  {Physical Review B}\ }\textbf {\bibinfo {volume} {103}},\ \href
  {https://doi.org/10.1103/PhysRevB.103.094109} {10.1103/PhysRevB.103.094109}
  (\bibinfo {year} {2021})\BibitemShut {NoStop}%
\bibitem [{\citenamefont {Pizzi}\ \emph {et~al.}(2022)\citenamefont {Pizzi},
  \citenamefont {Malz}, \citenamefont {Nunnenkamp},\ and\ \citenamefont
  {Knolle}}]{Pizzi_2022}%
  \BibitemOpen
  \bibfield  {author} {\bibinfo {author} {\bibfnamefont {A.}~\bibnamefont
  {Pizzi}}, \bibinfo {author} {\bibfnamefont {D.}~\bibnamefont {Malz}},
  \bibinfo {author} {\bibfnamefont {A.}~\bibnamefont {Nunnenkamp}},\ and\
  \bibinfo {author} {\bibfnamefont {J.}~\bibnamefont {Knolle}},\ }\bibfield
  {journal} {\bibinfo  {journal} {Physical Review B}\ }\textbf {\bibinfo
  {volume} {106}},\ \href {https://doi.org/10.1103/PhysRevB.106.214303}
  {10.1103/PhysRevB.106.214303} (\bibinfo {year} {2022})\BibitemShut {NoStop}%
\bibitem [{\citenamefont {Richter}\ \emph {et~al.}(2023)\citenamefont
  {Richter}, \citenamefont {Lunt},\ and\ \citenamefont
  {Pal}}]{Richter:2022clf}%
  \BibitemOpen
  \bibfield  {author} {\bibinfo {author} {\bibfnamefont {J.}~\bibnamefont
  {Richter}}, \bibinfo {author} {\bibfnamefont {O.}~\bibnamefont {Lunt}},\ and\
  \bibinfo {author} {\bibfnamefont {A.}~\bibnamefont {Pal}},\ }\href
  {https://doi.org/10.1103/PhysRevResearch.5.L012031} {\bibfield  {journal}
  {\bibinfo  {journal} {Phys. Rev. Res.}\ }\textbf {\bibinfo {volume} {5}},\
  \bibinfo {pages} {L012031} (\bibinfo {year} {2023})},\ \Eprint
  {https://arxiv.org/abs/2205.06309} {arXiv:2205.06309 [cond-mat.stat-mech]}
  \BibitemShut {NoStop}%
\bibitem [{\citenamefont {Roberts}\ and\ \citenamefont
  {Swingle}(2016)}]{Roberts:2016wdl}%
  \BibitemOpen
  \bibfield  {author} {\bibinfo {author} {\bibfnamefont {D.~A.}\ \bibnamefont
  {Roberts}}\ and\ \bibinfo {author} {\bibfnamefont {B.}~\bibnamefont
  {Swingle}},\ }\href {https://doi.org/10.1103/PhysRevLett.117.091602}
  {\bibfield  {journal} {\bibinfo  {journal} {Phys. Rev. Lett.}\ }\textbf
  {\bibinfo {volume} {117}},\ \bibinfo {pages} {091602} (\bibinfo {year}
  {2016})},\ \Eprint {https://arxiv.org/abs/1603.09298} {arXiv:1603.09298
  [hep-th]} \BibitemShut {NoStop}%
\bibitem [{\citenamefont {Lieb}\ and\ \citenamefont
  {Robinson}(1972)}]{Lieb:1972wy}%
  \BibitemOpen
  \bibfield  {author} {\bibinfo {author} {\bibfnamefont {E.~H.}\ \bibnamefont
  {Lieb}}\ and\ \bibinfo {author} {\bibfnamefont {D.~W.}\ \bibnamefont
  {Robinson}},\ }\href {https://doi.org/10.1007/BF01645779} {\bibfield
  {journal} {\bibinfo  {journal} {Commun. Math. Phys.}\ }\textbf {\bibinfo
  {volume} {28}},\ \bibinfo {pages} {251} (\bibinfo {year} {1972})}\BibitemShut
  {NoStop}%
\bibitem [{\citenamefont {Nachtergaele}\ \emph {et~al.}(2006)\citenamefont
  {Nachtergaele}, \citenamefont {Ogata},\ and\ \citenamefont
  {Sims}}]{Nachtergaele_2006}%
  \BibitemOpen
  \bibfield  {author} {\bibinfo {author} {\bibfnamefont {B.}~\bibnamefont
  {Nachtergaele}}, \bibinfo {author} {\bibfnamefont {Y.}~\bibnamefont
  {Ogata}},\ and\ \bibinfo {author} {\bibfnamefont {R.}~\bibnamefont {Sims}},\
  }\href {https://doi.org/10.1007/s10955-006-9143-6} {\bibfield  {journal}
  {\bibinfo  {journal} {Journal of Statistical Physics}\ }\textbf {\bibinfo
  {volume} {124}},\ \bibinfo {pages} {1} (\bibinfo {year} {2006})}\BibitemShut
  {NoStop}%
\bibitem [{\citenamefont {Hastings}\ and\ \citenamefont
  {Koma}(2006)}]{Hastings_2006}%
  \BibitemOpen
  \bibfield  {author} {\bibinfo {author} {\bibfnamefont {M.~B.}\ \bibnamefont
  {Hastings}}\ and\ \bibinfo {author} {\bibfnamefont {T.}~\bibnamefont
  {Koma}},\ }\href {https://doi.org/10.1007/s00220-006-0030-4} {\bibfield
  {journal} {\bibinfo  {journal} {Communications in Mathematical Physics}\
  }\textbf {\bibinfo {volume} {265}},\ \bibinfo {pages} {781} (\bibinfo {year}
  {2006})}\BibitemShut {NoStop}%
\bibitem [{\citenamefont {Hastings}(2010)}]{Hastings:2010lrb}%
  \BibitemOpen
  \bibfield  {author} {\bibinfo {author} {\bibfnamefont {M.~B.}\ \bibnamefont
  {Hastings}},\ }\href {https://arxiv.org/abs/1008.5137} {\  (\bibinfo {year}
  {2010})},\ \Eprint {https://arxiv.org/abs/1008.5137} {arXiv:1008.5137
  [math-ph]} \BibitemShut {NoStop}%
\end{thebibliography}%
\end{document}